\documentclass{emulateapj}
                             
\shorttitle{Integral field spectroscopy of Arp~299}
\shortauthors{M. Garc\'{\i}a-Mar\'{\i}n et al.}

\begin{document}

\title{Integral field spectroscopy of the Luminous Infrared Galaxy Arp~299 (IC~694+NGC~3690)\altaffilmark{1}}

\author{M. Garc\'{\i}a-Mar\'{\i}n\altaffilmark{2}, L. Colina\altaffilmark{2},
  S. Arribas\altaffilmark{3,4,5},
  A. Alonso-Herrero\altaffilmark{2}, E. Mediavilla\altaffilmark{5}}
\altaffiltext{1}{Based on observations with the William Herschel Telescope
  operated on the island of La Palma by the ING in the Spanish Observatorio
  del Roque de los Muchachos of the Instituto de Astrof\'{\i}sica de
  Canarias. Based also on observations with the NASA-ESA Hubble Space
  Telescope, obtained at the Space Telescope and Science Institute, which is
  operated by the Association of Universities for Research in Astronomy,
  Inc. under NASA contract number NAS5-26555.}
\altaffiltext{2}{Departamento de Astrof\'{\i}sica Molecular 
e Infrarroja, Instituto de Estructura de la Materia, CSIC, E-28006 Madrid, Spain.}
\altaffiltext{3}{Space Telescope Science Institute, 3700 San Martin Drive, Baltimore, M.D. 21218, USA.}
\altaffiltext{4}{On leave from the Instituto de Astrof\'{\i}sica de Canarias (IAC) and from the Consejo Superior de Investigaciones Cient\'{\i}ficas (CSIC), Spain.}
\altaffiltext{5}{Instituto de Astrof\'{\i}sica de Canarias (IAC), 38205 La
  Laguna, Canary Islands, Spain.}

\begin{abstract}
The luminous infrared galaxy Arp~299 (IC~694+NGC~3690) is studied using optical
integral field spectroscopy obtained with the INTEGRAL system, 
together with archival
\textit{Hubble Space Telescope} WFPC2 and NICMOS images. The
stellar and ionized gas morphology shows $\lambda$-dependent variations due to
the combined effects of the dust internal extinction, and the nature and
spatial distribution of the different ionizing sources. The two-dimensional
ionization maps have revealed an off-nuclear conical structure
of about 4 kpc in length, characterized by high excitation conditions and a radial
gradient in the gas electron density. The apex of this structure
coincides with B1 region of NGC~3690 which, in turn, presents Seyfert-like
ionization, high extinction and a high velocity dispersion. These
results strongly support the hypothesis that B1 is the true nucleus of NGC~3690, where an AGN is located. In the circumnuclear regions
H\,{\sc ii}-like ionization dominates, while LINER-like ionization is found elsewhere. 
The H$\alpha$ emitting sources with ages from 3.3 to 7.2$\times$10$^{6}$ 
years, have masses of
between 6 and 680$\times$10$^{6}$ M$\odot$ and contribute (extinction corrected) 
about 45$\%$ to the bolometric luminosity. The ionized (H$\alpha$) and neutral (NaD) gas velocity
fields show similar structure on scales of several hundred to
 about 1 kpc, indicating that these gas components are kinematically coupled. 
The kinematic structure is complex and on scales of about 0.2 kpc does not appear 
to be dominated by the presence of ordered, rotational motions. The large
velocity dispersion measured in NGC~3690 indicates that this galaxy is the
most massive of the system. The low velocity amplitude and dispersion
of the interface suggest that the ionized gas is slowly rotating or in
a close to quiescent phase.
\end{abstract}

\section{Introduction}

Arp~299 is a nearby luminous infrared galaxy  (d$\sim$43 Mpc for H$_{0}$=70 km~s$^{-1}$Mpc$^{-1}$, $\Omega_{M}$= 0.7, $\Omega_{\Lambda}$= 0.3) with  an
infrared (8-1000 $\mu$m) luminosity of 
L$_{IR}$=5.7$\times$10$^{11}$L$_{\odot}$\footnote{IRAS fluxes obtained from
  Moshir et al. (1993). L$_{IR}$ derived from the equation detailed on Sanders
  \& Mirabel (1996).}.  It  is an interacting system (Fig. \ref{galaxia})
composed of two individual galaxies (IC~694+NGC~3690\footnote{Through this paper, we use the
  notation introduced by Gehrz et al (1983) for the two components of Arp~299: the nucleus of IC~694 (eastern component) is region
  A, while the main sources in NGC~3690 (western component) are B1, B2,
  C(C1+C2+C3) and C$'$ (see also Alonso-Herrero et al 2000).}) in an early dynamical stage. Hibbard \& Yun (1999) have
discovered an H\,{\sc i} tidal tail of 180 kpc in length, with no evidence for tidal dwarf galaxies.  Powerful starburst regions with star formation rates of up to about 100 M$_{\odot}$ yr$^{-1}$ have been identified (Alonso-Herrero et al. 2000, hereafter AAH00). Moreover, the
system presents several regions where supernova remnants have been detected at
radio wavelengths (Neff et al.
2004). More recently, Mattila et al. (2005) have discovered a new
supernova using near-IR images, reinforcing the prediction that this galaxy has a high rate of
supernova explosions (Mannucci et al. 2003).

Optical spectroscopic studies have classified IC~694 as starburst, and NGC~3690 as starburst/LINER (e.g., Coziol et al. 1998), and mid-IR
observations classified the system as starburst (e.g., Laurent et al. 2000). Using
\textit{Hubble Space Telescope} (HST) imaging and near-infrared spectroscopy, AAH00 modeled the star formation
properties of the system without the need for an AGN. The first hint for the presence of an AGN in Arp~299 came from
X-ray observations with the \textit{BeppoSAX} satellite
by Della Ceca et al. (2002). These authors concluded that a 
highly absorbed (N$_{H}$~$\simeq$~2.5$\times$10$^{24}$ cm$^{-2}$) AGN,
with an intrinsic luminosity of L$_{0.5-100 keV}$ $\simeq$ 1.9$\times$10$^{43}$ 
erg s$^{-1}$, is present in this galaxy, although due to the low spatial
resolution its exact location could not be determined. Using \textit{Chandra}
and \textit{XMM-Newton}
observations, recent studies (Ballo et al. 2004) have inferred the presence of
two AGNs, one located in NGC~3690 (source B1; 6.4 keV line), and another in 
IC~694 (source A; 6.7 keV Fe-H$\alpha$ line). In addition, Gallais et al. (2004) 
suggest the presence of an AGN in NGC~3690 based on the shape of the mid-IR
continuum, while in IC~694 they only detect spectral features characteristic of 
massive, dust enshrouded star formation.

This paper presents a detailed
study of the two-dimensional star formation, kinematics, and ionized gas
properties of Arp~299, covering the two individual galaxies of the system and the
interface area. At a distance of about 43 Mpc, Arp~299 is one of the nearest
interacting, luminous infrared galaxies, and therefore these observations
provide high spatial resolution (about 0.205 kpc arcsec$^{-1}$) and allow us to investigate
in detail the structure and physical
properties of this galaxy. The paper is organized as follows: \S 2 describes the
observations and data reduction, while the selection of regions under study is
described in \S 3. The next sections present a discussion of the results, with special emphasis on the stellar and ionized gas morphology (\S
\ref{sec4}), the internal dust distribution and extinction effects (\S \ref{sec5}), the extinction
estimates using optical and near-IR emission lines (\S \ref{sec54}), the two
dimensional ionization structure (\S \ref{sec6}), the electron density (\S
\ref{electron}), the origin of the LINER-like ionization (\S \ref{sec8}), the
star-forming properties (\S \ref{sec9}), and the kinematics and dynamical mass
derivations (\S \ref{sec-kin}). Finally \S \ref{sec11} presents a short summary of the main new results.

\section{Observations and Data Reduction} 
\begin{figure*}
\epsscale{1.}
\plotone{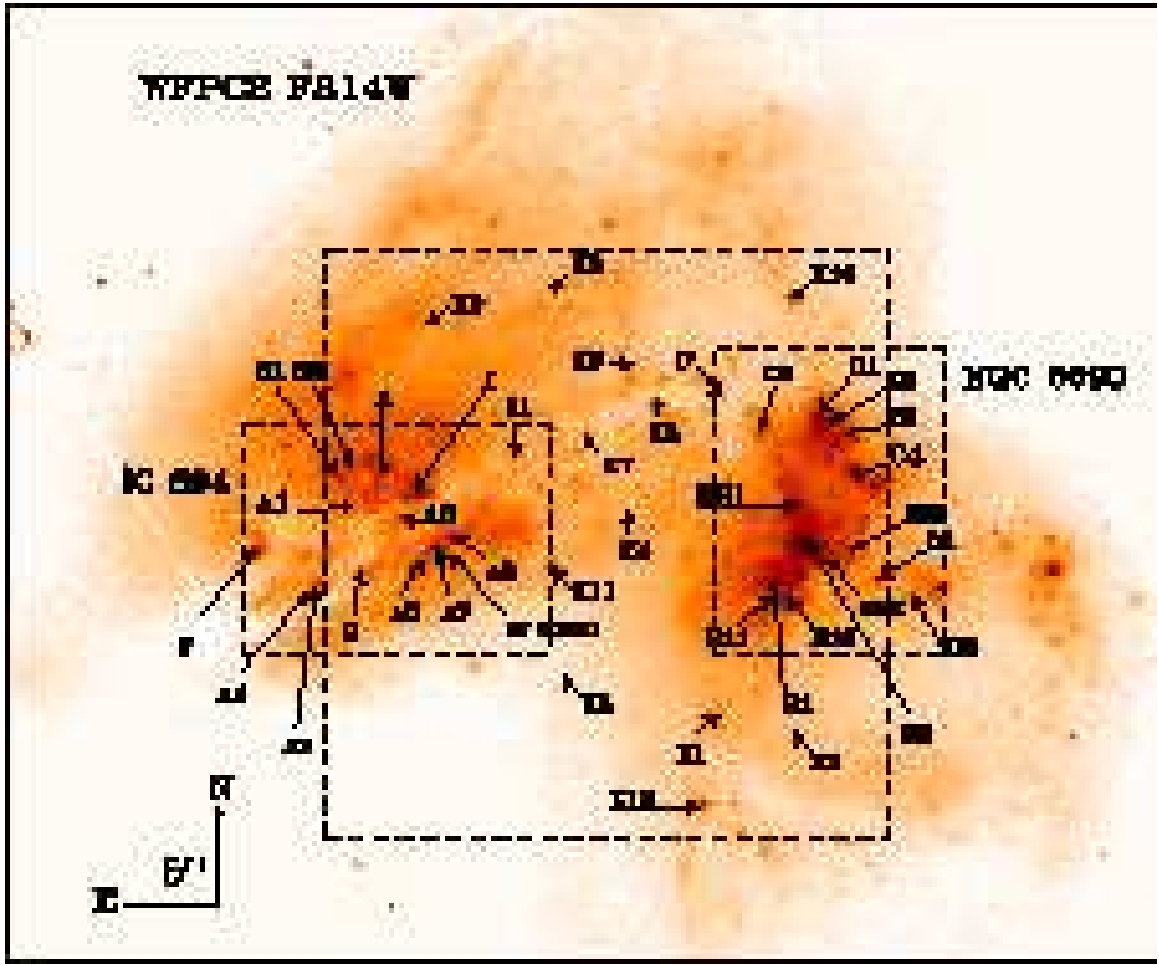}
\caption{Archival HST image of the interacting system Arp~299 obtained with
  the WFPC2 F814W filter. The rectangular overlays indicate the
  effective area covered by the three pointings with the two INTEGRAL
  configurations used (i.e SB2 small rectangles, SB3 large rectangle). We have
  marked the positions of the regions under study (see \S\ 3 and Tables 1, 2,
  and 3). The location of the recently discovered near-IR SN is shown too
  (Mattila et al. 2005). The image is shown on a logarithmic
  scale. [\textit{See the electronic edition of the Journal for a color version of this figure.}]}\label{galaxia}
\end{figure*}

Integral Field Spectroscopy (IFS) of Arp~299 was obtained on 
2004 January 15 and 17 using INTEGRAL, a fiber-based integral field
system (Arribas et al. 1998) connected to the Wide-Field Fiber Optic
Spectrograph (WYFFOS) (Bingham et al. 1994) and mounted on the 4.2 m William Herschel
Telescope. Two different INTEGRAL configurations were used
for observing the system (see pointings in Fig. \ref{galaxia}).
The fiber bundle with the largest field of view (FoV), SB3, was centered on a
position equidistant from the centers of the galaxies and therefore used to map the
diffuse, interface region between the two galaxies (IC~694+NGC~3690) that form the Arp~299
system, as well as the 
galaxy external regions at distances of about 4 kpc. The SB3 bundle
 consists of 115 fibers each 2\farcs7
  in diameter, covering a rectangular area of 33\farcs6$\times$29\farcs4 on the
  sky. In addition 20 fibers, forming a circle of 45\farcs0 in radius, were
  used to obtain a spectrum of the sky simultaneously. 
The SB2 bundle was used to map
each of the two galaxies (IC~694 and NGC~3690, eastern and western galaxies,
respectively on Fig. \ref{galaxia}) individually. It consists
 of 189 fibers each 0\farcs9
  in diameter, covering a rectangular area of 16\farcs0$\times$12\farcs3 on the
  sky. In addition 30 fibers, forming a circle of 45\farcs0 in radius, were
  used to measure the sky simultaneously.  

The spectra were taken using a 600 lines mm$^{-1}$ grating, with an effective spectral
resolution (FWHM) of approximately 6.0 \AA\ and 9.8 \AA\ for the SB2 and SB3 bundles
respectively, and covering the spectral range $\lambda\lambda$4400$-$7000 \AA .
The total integration time was 3600 s for each pointing with the SB2 bundle, split into
three individual exposures  of 1200 s each. The total integration time with
the SB3 bundle was 5400 s divided into three individual exposures of 1800 s
each. The seeing conditions were fairly stable during the
two periods of  observation (about 0\farcs7 the first night, when we used the SB2
bundle,  and 1\farcs1 the second night when we used the SB3 bundle).

The data reduction consists of three main steps: 1) the basic reduction of two
dimensional fiber spectra, 2) the line fitting, and 3) the generation of maps
of spectral features (such as intensity maps and velocity fields) from the
calibrated spectra. The basic reduction was performed using the IRAF\footnote{\texttt{IRAF} software is distributed by the \textit{National Optical Astronomy Observatories (NOAO)} which is operated by the \textit{Association of Universities for Research in Astronomy, Inc. (AURA)} in cooperation with the \textit{National Science Foundation}.} environment, and followed the standard procedures applied to spectra obtained with two-dimensional fiber spectrographs (Arribas et al. 1997 and references therein). The absolute flux calibration was performed observing the standard star
   Feige~34 with the same instrument configuration and data reduction procedures 
   used on the galaxy. For each spectrum, the total flux, the radial
   velocities and the velocity dispersions ($\sigma$) were measured by fitting single
   Gaussian functions to the observed emission line profiles using the DIPSO
   package (Howarth \& Murray 1988). For the generation of the maps, a two dimensional
  interpolation was applied using the free software IDA
 (Garc\'{\i}a-Lorenzo et al. 2002). The blue and red stellar continuum images
  were taken as an emission line-free filter centered at $\lambda_{restframe}$=4475 \AA\ and
  6525 \AA\ respectively, and with a rectangular bandwidth of 150 \AA\,. The measures on the
  SB3 maps were done through an aperture of 3\farcs0 in diameter, while
  for the SB2 maps the apertures diameter were 1\farcs0 (IC~694) and
  0\farcs9 (NGC~3690, to avoid the overlap between regions).

\begin{figure*}
\epsscale{0.8}
\plotone{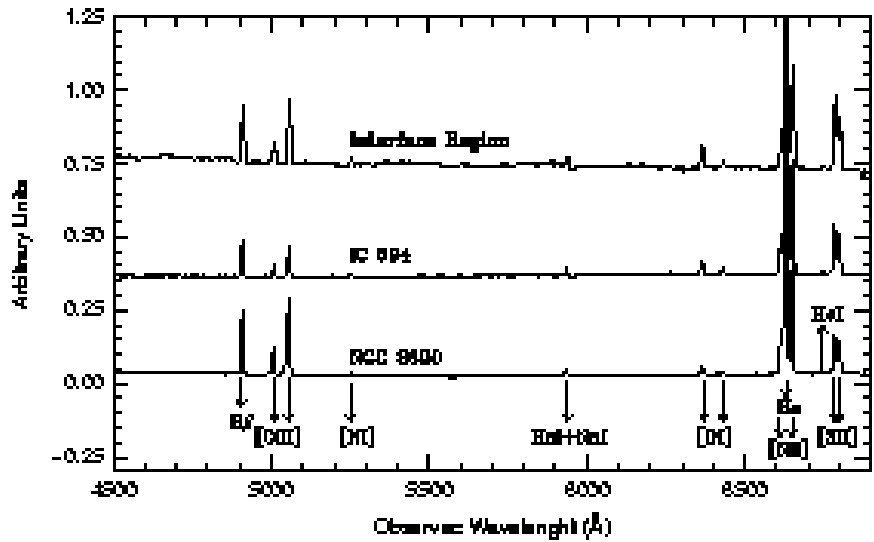}
\caption{Average spectra of the Interface Region (K1 to K12), IC~694 (A, H1 and H2) and NGC~3690 (B1 and B11). The spectral range shown is 
4500-6900\AA. See Fig. \ref{galaxia} and Tables \ref{infoarp}, \ref{tabic},
and \ref{tabngc} for location of the individual emitting sources.}\label{espectrillos}
\end{figure*}

Three sample spectra, covering the entire used spectral range, are shown in
Fig. \ref{espectrillos}.  All the
spectra show the strong (H$\beta$, [O\,{\sc
    iii}]$\lambda\lambda$4959, 5007, 
    H$\alpha$+[N\,{\sc ii}]$\lambda\lambda$6548, 6584, 
    He\,{\sc i}$\lambda$6678, and [S\,{\sc ii}]$\lambda\lambda$6716, 6731) 
    and the weak (N\,{\sc i}$\lambda$5199, He\,{\sc i}$\lambda$5876,
    [O\,{\sc i}]$\lambda\lambda$6300, 6364, and He\,{\sc i}$\lambda$6678)
     emission lines in different ratios, as well as interstellar absorption lines
      (Na\,{\sc i}$\lambda\lambda$5890, 5896), that will be used to trace the
     neutral interstellar medium kinematics.

Complementary HST multiwavelength images both in continuum and emission lines
are also available. These additional data include  the red 
(WFPC2-F814W, archive images) and the near-infrared (NICMOS1-F110M,
NICMOS2-F160W and NICMOS2-F222M, from AAH00) continuum images. These filters are the HST
equivalent to the ground-based Johnson-Cousins system \textit{I}, \textit{J},
\textit{H} and \textit{K} respectively (Origlia \& Leitherer 2000). Images of
the ionized gas as traced by the Pa$\alpha$ ($\lambda_{rest}$=1.875 $\mu$m) light (NICMOS-F190N, from AAH00)
will be used too. The archival WFPC2 image was calibrated on-the-fly with the
best available reference files, while the NICMOS images were reduced with
routines from the package NicRed (McLeod 1997; see details on AAH00).  

\section{Selection of regions under study}
In order to analyze the system we have selected several regions distributed
over the individual galaxies, the interface between them and regions external
to the galaxies (Fig. \ref{galaxia}). The selection includes compact
unresolved high-surface brightness knots, star clusters and more diffuse, extended resolved low-surface brightness areas. For IC~694 and
NGC~3690 we based our selection on that done by AAH00, that includes the nuclear
regions (A and B1 respectively) and the most important star-forming regions and bright star clusters.
To identify the position of the regions under study in the lower resolution
 INTEGRAL maps, a direct comparison of the convolved WFPC2 image with the red
 continuum image was performed. The following sections present the main results obtained for these different regions and discuss their implications separately for the two individual galaxies, 
IC~694 and NGC~3690, and for the interface region between them. The physical
properties measured for each individual region are presented in Tables 1 to 3. 

\section{Stellar and Ionized Gas Morphology}\label{sec4}
Arp~299 is the result of a merger between the galaxies IC~694 and NGC~3690, whose nuclei (A and B1 respectively,
see Fig. \ref{interfaceinfo}) are located 
 about 4.6~kpc apart. The system shows a morphologically disturbed 
 structure characterized by the presence of compact high-surface
 knots, and a complex system of filaments and dust lanes, as already indicated by previous
studies at several wavelengths (e.g. Gerhz et al. 1983; Telesco et al. 1985;
Baan et al. 1990; AAH00). In what follows, we discuss the differences
in the spatial distribution of the stellar and ionized gas components.  

\subsection{Interface Region}

\begin{figure*}
\epsscale{1}
\plotone{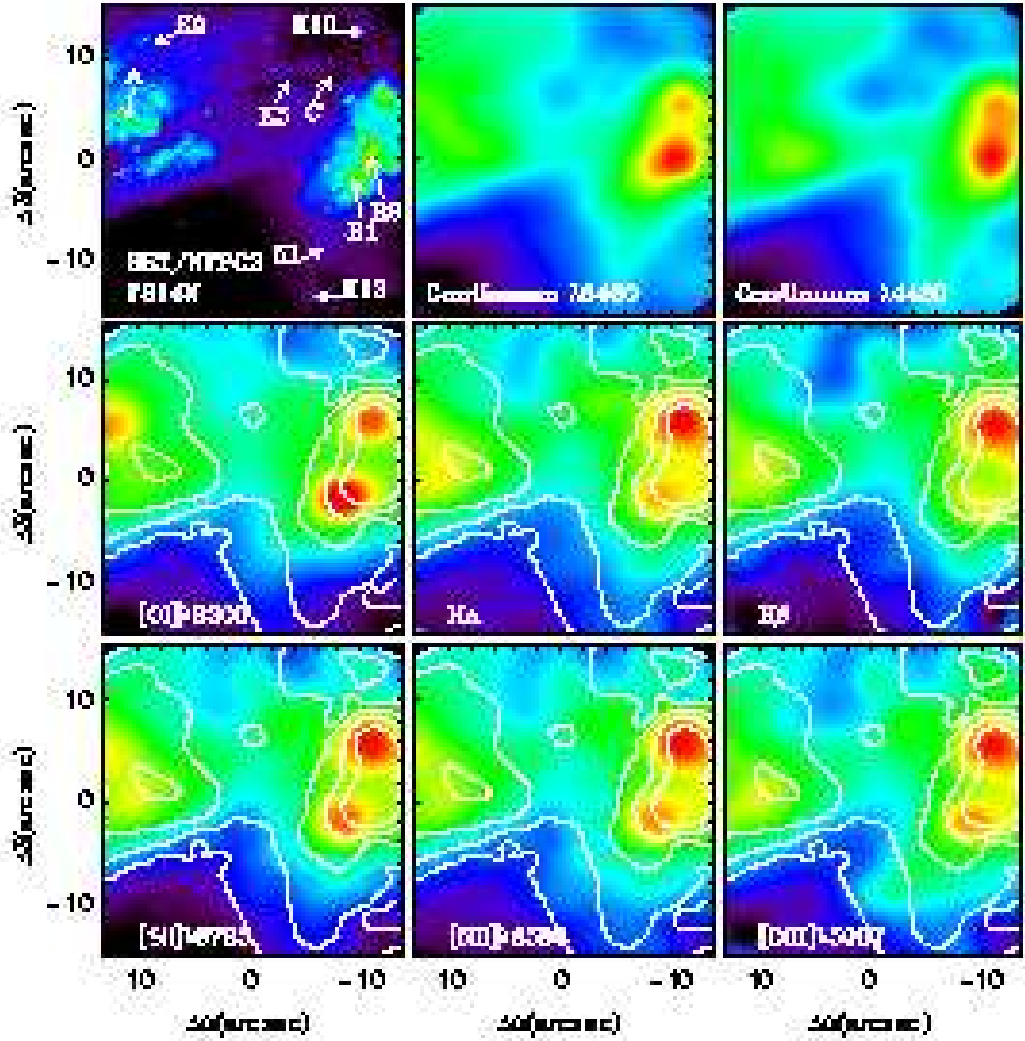}
\caption{Emission-line free stellar continuum and emission line images for Arp~299 obtained with INTEGRAL SB3 bundle. NGC~3690 (B1 and B2) is
  located to the west and IC~694 (A) to the east.
 The contours represent the red continuum at $\lambda$6460. The HST /WFPC2
  F814W image is shown for comparison, where several regions of interest
discussed in the text are marked. All the images are shown on a logarithmic
  scale, and the color code represents different intensity levels in
  each of the individual
  maps. Orientation is north up, east to the left. [\textit{See the electronic edition of the Journal for a color version of this figure.}]}\label{interfaceinfo}
\end{figure*}

 The two INTEGRAL/SB3 continuum images of Arp~299, obtained at wavelengths
 blueward of H$\alpha$ and
 H$\beta$, reproduce well the overall stellar morphology 
 detected with the HST WFPC2 F814W image (Fig. \ref{interfaceinfo}) although
 at a lower spatial resolution (fiber diameter 2\farcs7, scales of $\simeq$0.5 kpc). In spite of that, the interface
 region between the two galaxies, and the external regions
 show well defined features such as the bridge to the south
 of region K5 connecting NGC~3690 and IC~694, and the elongated emission
 connecting K1 towards K12 source as part of a tail-like structure
 located to the SW of NGC~3690 (see Fig. \ref{galaxia}). 

 The general structure of the low surface brightness ionized gas in the interface region, as traced by for example H$\alpha$
 (Fig. \ref{interfaceinfo}), is similar to that of the stellar
 component. However, several differences exist.
 Diffuse ionized gas is detected NE of NGC~3690, associated
 with the K5 and C$'$ knots; the latter region shows Pa$\alpha$ emission and SN
 activity in radio (Neff et al. 2004), and is part of a bridge of ionized
 material (see Fig. \ref{interfaceinfo}, H$\alpha$ map) to the north of the 
 stellar bridge that connects the two individual galaxies. The
 [O\,{\sc{iii}}]$\lambda$5007 emission map presents an elongated structure
 located towards the south-east of B1 and B2 tracing an area where the ionized
 gas presents high excitation state (see \S \ref{ionizinterface}). However, as 
 showed in the next two sections, the most important morphological variations are
 associated to the main bodies of the galaxies when observed with the higher
 angular resolution SB2 bundle.

\subsection{IC~694}

\begin{figure*}
\epsscale{1.0}
\plotone{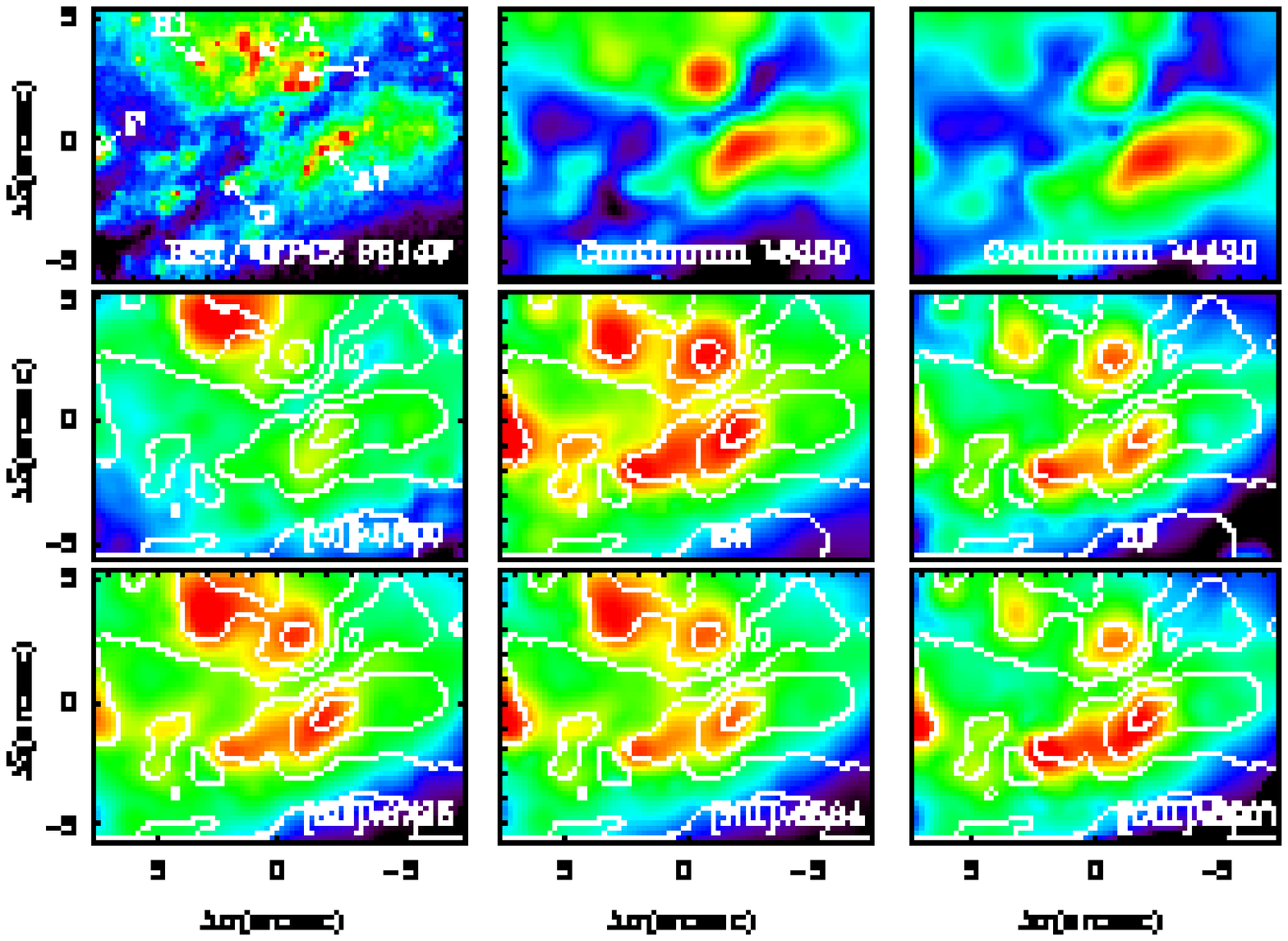}
\caption{Emission-line free stellar continuum and emission line images for IC~694 obtained with
 INTEGRAL SB2 bundle. The contours represent the red continuum at $\lambda$6460. The HST/WFPC2 F814W image is also
 shown for comparison, with several regions of interest marked on it.  All the images are shown on a logarithmic
  scale, and the color code represents different intensity levels in
  each of the individual maps. Orientation is north up, east to the left. [\textit{See the electronic edition of the Journal for a color version of this figure.}]}\label{icinfo}
\end{figure*}

 Previous studies have concluded that IC~694 was a spiral galaxy
that due to the merging process shows a more disrupted morphology 
(e.g., Gerhz et al. 1983; Standford \& Wood 1989; Hibbard \& Yun 1999).  The 
stellar structure as seen from both the HST and the INTEGRAL continuum images
(see Fig. \ref{icinfo}) is divided into two
main substructures: (1) The arm 
(region centered around A7) of about 1.8 kpc in length, that could be one of
the original spiral arms disrupted and deformed as a
consequence of the merging-process, and (2) the main body of the galaxy, formed by
the dust-enshrouded infrared nucleus A (AAH00), and surrounding regions. The two
structures are well separated (distances of $\simeq$1 kpc)
by what seems to be a dust lane, that is clearly visible in the HST
near-IR color maps (AAH00). The effect of the internal extinction is
remarkable in the continuum images
as the relative surface brightness of the regions, in particular nucleus A and
surrounding regions, changes with wavelength (see  Fig. \ref{icinfo} and
discussion in \S \ref{sec52}).

The ionized gas distribution (see Fig. \ref{icinfo}) shows several distinct 
features worth mentioning. The nucleus of the galaxy (A) appears as a
weak optical line emitting source, whereas the regions of high surface brightness are located
at different positions (e.g., part of the arm and towards east, and regions
close to A) and separated up to 1 kpc from
the nucleus. However, in the near-IR A is the brightest Pa$\alpha$ and Br$\gamma$
source (see AAH00 and Sugai et al. 1999, see also \S \ref{sec52}),
indicating that the extinction towards the nucleus is very high (using
Pa$\alpha$/H$\alpha$ ratio, A$_{V}$=6.1$\pm$0.7). There are regions (e.g. knot
G) that appear as a faint continuum
sources, but that are clearly identified as strong optical line emitters.  In addition to
differences in the morphology of the ionized gas (i.e. excitation conditions) with respect to that of the 
continuum, there are also differences in the ionized gas distribution as traced
by several lines. The hydrogen recombination lines and the  [N\,{\sc ii}]$\lambda$6584 and [S\,{\sc ii}]$\lambda\lambda$6717, 6731 
emission lines show a similar distribution, dominated by the emission from 
regions located on the arm and towards the east, and in the regions
surrounding the nucleus.
However, the shock tracer [O\,{\sc i}]$\lambda$6300 line is concentrated in
region H1, while the higher excitation line [O\,{\sc iii}]$\lambda$5007
appears to be brightest in regions A7 and G. These differences in the distribution of the emission line gas are not only due 
to internal extinction effects, but they also reflect changes in the ionization conditions
as a function of location within the galaxy (see \S \ref{ionizinterface}).

 \subsection{NGC~3690}
\begin{figure*}
\epsscale{0.8}
\plotone{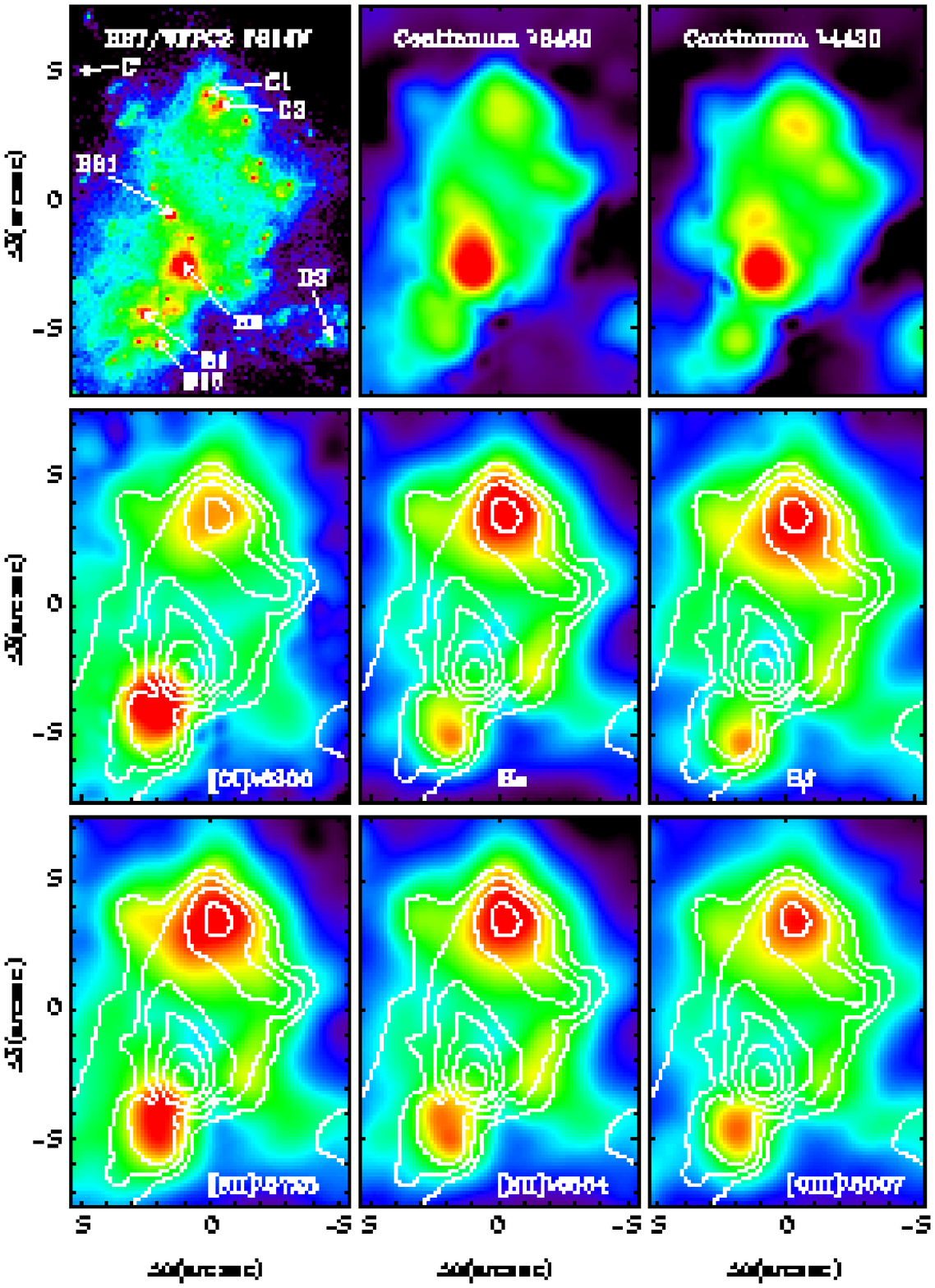}
\caption{Emission-line free stellar continuum and emission line images for NGC~3690 obtained with
 INTEGRAL SB2 bundle. The contours represent the red continuum at $\lambda$6460. The HST/WFPC2 F814W image is also
 shown for comparison, with several regions of interest marked on it. All the images are shown on a logarithmic
  scale, and the color code represents different intensity levels in
  each of the individual maps. Orientation is north up, east to the left. [\textit{See the electronic edition of the Journal for a color version of this figure.}]}\label{ngcinfo}
\end{figure*}

The stellar light distribution as observed with the HST shows an irregular
 morphology dominated by the  emission coming from the high surface brightness
 region B2 (identified therefore as the optical nucleus), and from region B1
 and complex C\footnote{As explained in Note 7, we will use the term
 complex  C introduced by Gehrz et al. (1983) for the regions C1, C2 and C3
 (see Fig. \ref{galaxia}).}
 located at distances of 0.49  and 1.3 kpc from B2, respectively (see Fig.
  \ref{ngcinfo}). However at near and mid-IR wavelengths the true nucleus of
 the galaxy (see discussion in \S\ref{truenucleus}), B1, becomes the brightest continuum region in NGC~3690 (see
 AAH00; Satyapal et al. 1999). This
 disordered morphology could be the combination of  the 
  original spiral structure  with young  star-forming regions such as
  the complex C, and regions D3 or C$'$, newly formed as a consequence 
  of the interaction process as suggested by several authors  
 (e.g. Gerhz et al. 1983; Standford \& Wood 1989; Baan \& Haschick 1990;
 Hibbard \& Yun 1999). 

 The distribution of the ionized gas (Fig. \ref{ngcinfo}) is simpler than that
 of IC~694. The optical nucleus (B2) appears as an extremely faint line
 emitter, and the structure of the ionized gas is dominated by two high
 surface brightness regions separated by 1.8 kpc, nucleus B1 and the complex C (with the peak located
 towards C2), as previously  founded with near-IR hydrogen
 recombination lines (Sugai et al. 1999; AAH00). The brightest emission line 
 peak is associated to C2 for all lines but the shock tracer [O\,{\sc i}]$\lambda$6300. For the
 latter, the nucleus B1 is the brightest emitter, as also observed with the
 near-IR [Fe\, {\sc ii}]$\lambda$1.64 $\mu$m line (AAH00). Analyzing in detail
 the secondary emission line peaks associated to B1 and surrounding areas,
 it is interesting to note that the high excitation [O\,{\sc
 iii}]$\lambda5007$ emission is dominated by B1, while the
 H$\alpha$ and H$\beta$ emission peaks are associated to 
region B16 (which is located 0.26 kpc southwest of B1, and identified in the
 $\lambda$4475 continuum map). Low excitation collisional lines [N\,{\sc
 ii}]$\lambda$6584, [S\,{\sc ii}]$\lambda\lambda$6717, 6731 appear to have secondary peaks in both B1 and B16. These changes in the emission line distribution
represent changes in the ionization conditions, and most likely
in the ionizing source (AGN, young stellar clusters and shocks) on scales of less than 0.3 kpc.

\section{Internal dust distribution and extinction effects}\label{sec5}

\begin{figure*}
\epsscale{0.8}
\plotone{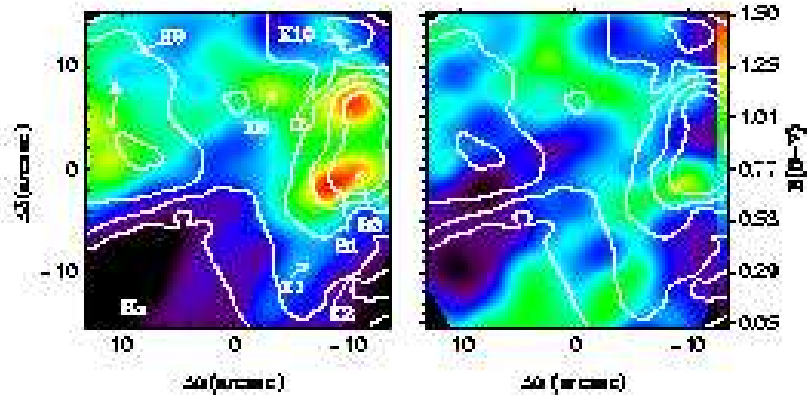}
\caption{\textit{Left panel:} Extinction corrected
  H$\alpha$ map of the Arp~299 system (including the interface region), shown on a logarithmic scale. \textit{Right panel:} Extinction map of the interface region obtained from
  the line ratio H$\alpha$/H$\beta$, shown on a linear scale. Contours are as in
  Fig. \ref{interfaceinfo}. [\textit{See the electronic edition of the Journal for a color version of this figure.}]}\label{ebvinterface}
\end{figure*}

As mentioned in the previous section, part of the complex structure observed
in the continuum and emission line light distributions is due to large amounts
of dust producing spatially dependent internal extinction effects. The two dimensional extinction maps of the gas (Figs. \ref{ebvinterface}, \ref{ebvic} and
\ref{ebvngc}) have been derived using the H$\alpha$/H$\beta$ line ratio, a
foreground dust screen model, and a mean interstellar extinction law based on
that of Savage \& Mathis (1979). No correction for the presence of underlying
stellar hydrogen absorption lines has been applied. Assuming an equivalent
width value of EW$_{abs}$(H$\alpha$) = EW$_{abs}$(H$\beta$) $\sim$ 2 \AA\ ,
the derived extinction could be overestimated by as much as $\Delta$E(B-V)
$\sim$ 0.6 for regions with extremely faint H$\beta$ emission
(e.g. EW(H$\beta$) $\leq$ 20~\AA). The values for the visual extinction
(A$_{V}$=3.1$\times$E(B-V)) in particular regions
are given in Tables \ref{infoarp}, \ref{tabic}, and \ref{tabngc}. The results suggest that the highest extinguished regions are, in general,
 associated to the nuclei and some particular regions (like C$'$), while the external H\,{\sc ii} regions are mostly less affected by the presence of
 dust. 

\subsection{Interface Region}
The extinction in the whole area covered by the SB3 bundle on scales of about 3\farcs0 (i.e. 0.6 kpc), is relatively uniform (see Fig. \ref{ebvinterface}). It has an average value
of A$_{V}$ = 1.9 mag on the selected individual regions, and ranges from
A$_{V}$(K2) = 0.9$\pm$0.7 mag to
A$_{V}$(K5) = 2.9$\pm$0.7 mag (see Table \ref{infoarp} for specific values on
all the individual regions). 

\subsection{IC~694}\label{sec52}
\begin{figure*}
\epsscale{0.8}
\plotone{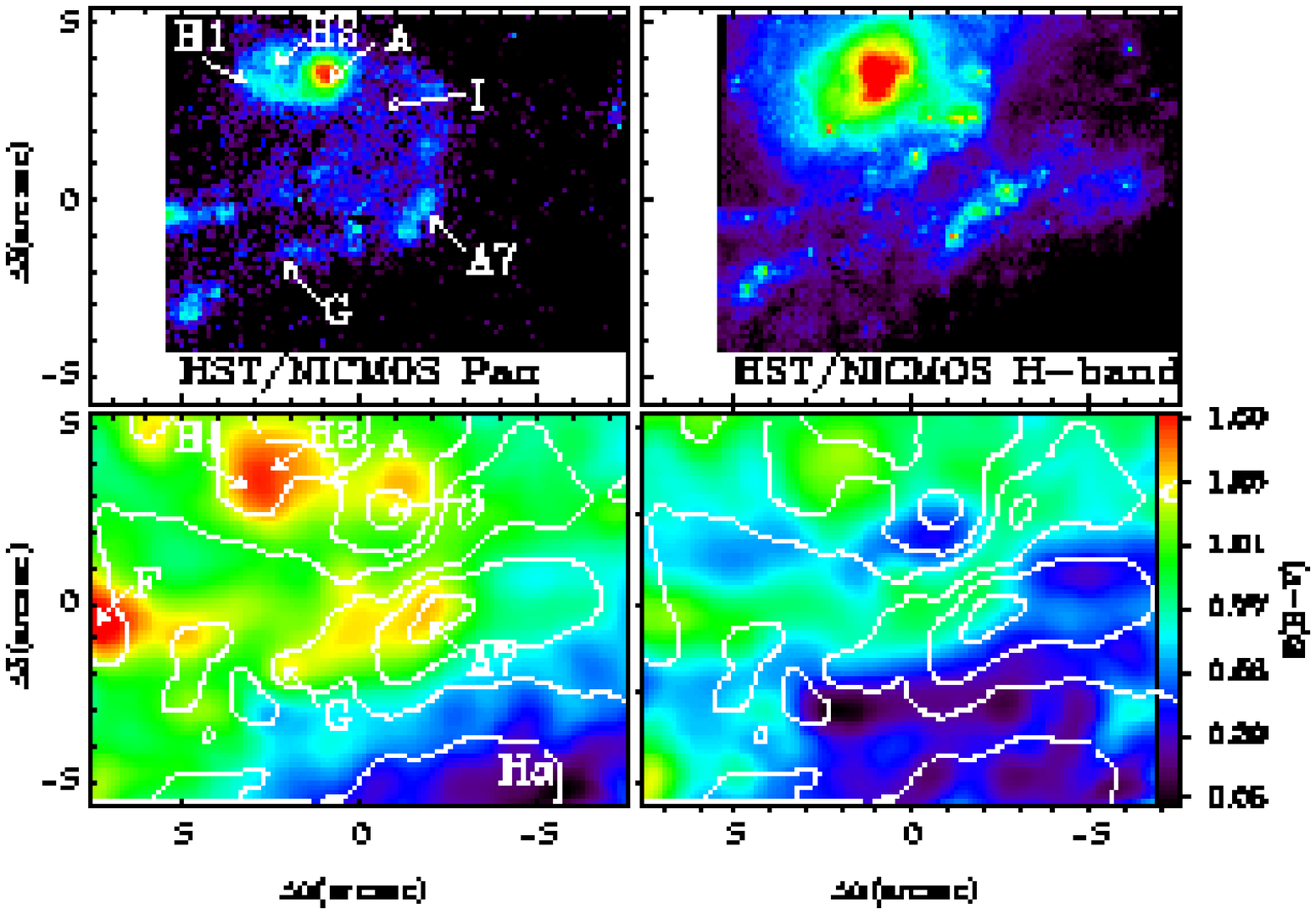}
\caption{IC~694. \textit{Top left panel:} HST/NICMOS continuum subtracted
 Pa$\alpha$ emission (F190N-F187N). \textit{Top right panel:} HST/NICMOS H-band continuum. \textit{Bottom left panel:} INTEGRAL H$\alpha$ extinction corrected map. \textit{Bottom right panel:}
 Extinction map obtained with H$\alpha$/H$\beta$ line ratio. As a
 reference several regions in the two fields are marked, and
 the red continuum contours are superimposed on the interpolated maps. The
 NICMOS images are shown covering the same FoV as the INTEGRAL
 data. All the images are shown on a logarithmic scale except
 for the extinction map, where a linear scale has been applied. [\textit{See the electronic edition of the Journal for a color version of this figure.}]}\label{ebvic}
\end{figure*}

On scales of  about 1\farcs0 (i.e. 0.21 kpc), the internal extinction in the visual covers a wide range of values from about
A$_{V}$=0.6$\pm$0.5 mag in the southwestern section of the
galaxy to A$_{V}$=3.4$\pm$0.5 mag in region H2, close to the nucleus A (see Table
\ref{tabic} for specific values, and Fig. \ref{ebvic} for the two dimensional E(B-V) distribution). As can be seen from Fig. \ref{ebvic}, the H$\alpha$
extinction corrected light distribution shows an overall structure similar to that of the
Pa$\alpha$ line except for the region associated with the peak of the
emission. While the nucleus A, identified as the brightest near-infrared source
in the HST images (Fig. \ref{ebvic}, see also AAH00), is also the dominant Pa$\alpha$ line emitter,
the peak of the extinction corrected H$\alpha$ distribution is located in
regions H1 and H2, 0.3 and 0.5 kpc towards the east of the nucleus. Also region F, located 1.58 kpc to the south-east of A,
appears as a strongly absorbed region (A$_{V}$=3.2$\pm$0.5 mag), and as a
strong H$\alpha$ emitter after correcting for extinction. Additional secondary
H$\alpha$ sources are mostly associated with star-forming regions (the arm, a
region towards the west of A) at distances of
1 to 1.6 kpc from the nucleus, and delineating what seems to be a spiral-like
arm structure previously identified in the Pa$\alpha$ emission line map.

\subsection{NGC~3690}\label{sec53}

\begin{figure*}
\epsscale{0.7}
\plotone{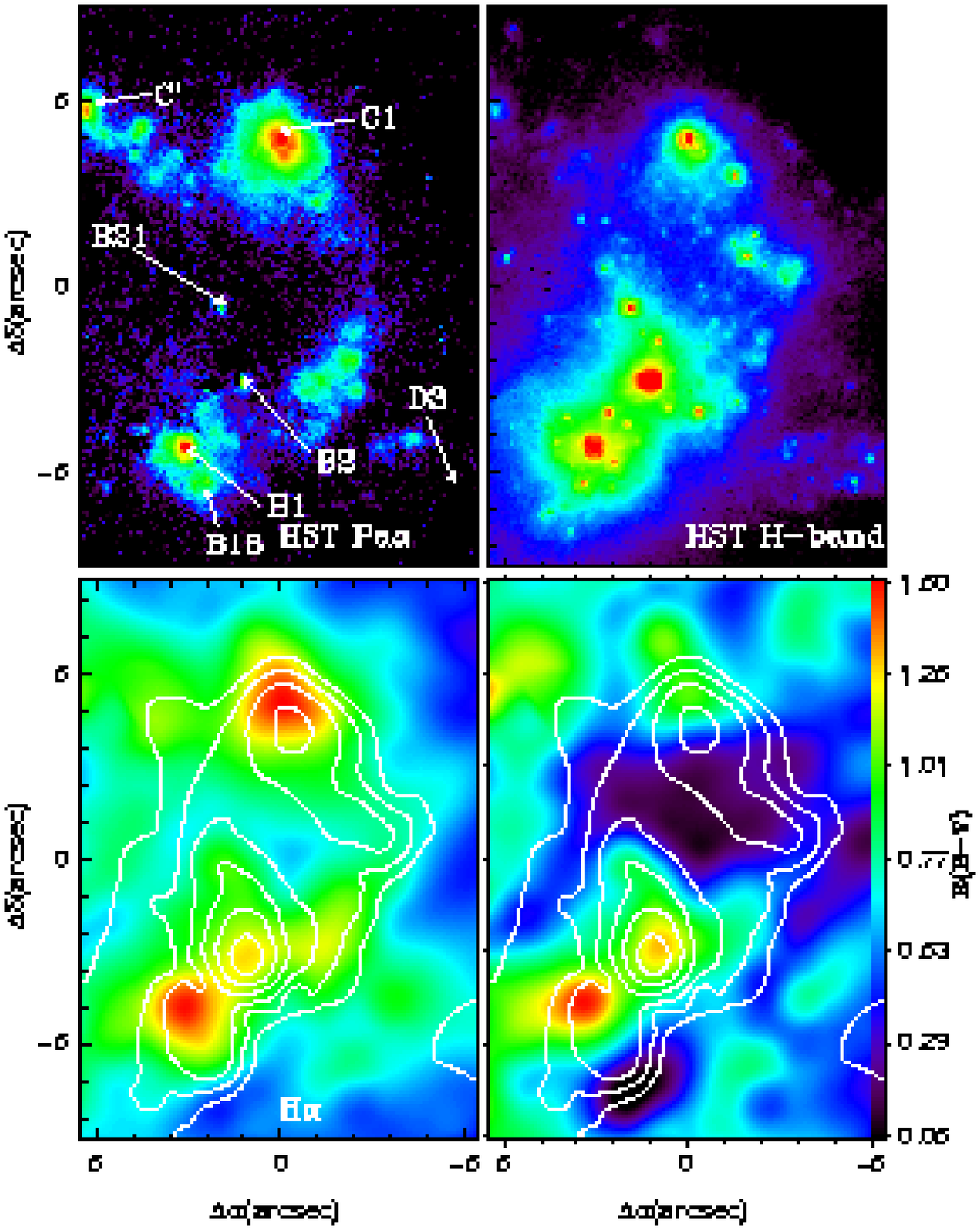}
\caption{NGC~3690. \textit{Top left panel:} HST/NICMOS continuum subtracted
 Pa$\alpha$ emission (F190N-F187N). \textit{Top right panel:} HST/NICMOS H-band continuum. \textit{Bottom
 left panel:} INTEGRAL H$\alpha$ extinction corrected map. \textit{Bottom right panel:}
 Extinction map obtained with H$\alpha$/H$\beta$ line ratio.  As a
 reference several regions are marked, and
 the red continuum contours are superimposed on the interpolated maps. The
 NICMOS images are shown covering the same FoV as the INTEGRAL
 data. All the images are shown on a logarithmic scale except
 for the extinction map, where a linear scale has been applied. [\textit{See the electronic edition of the Journal for a color version of this figure.}]}\label{ebvngc}
\end{figure*}

This galaxy presents the widest range of extinction (derived from the
H$\alpha$/H$\beta$ ratio) for the entire Arp~299
system with values that for the individual regions range from A$_{V}$(C4)=0.5$\pm$0.5 mag to
A$_{V}$(B2)=3.9$\pm$0.6 mag (see Table \ref{tabngc}). Also, the nucleus B1 in
this galaxy is the region of the Arp~299 system which has associated the highest H$\alpha$/H$\beta$-based extinction, A$_{V}\simeq$4.6 mag (Fig. \ref{ebvngc}).

The overall structure of the extinction corrected H$\alpha$ line 
distribution agrees well with that of the Pa$\alpha$ line (Fig. \ref{ebvngc},
see also AAH00) not only on the high surface brightness regions such as
the nucleus B1 and C1, but also on the low surface brightness region, including
 B2 identified as the optical nucleus. 

\subsection{Extinction estimate using optical and near-IR emission lines}\label{sec54}
Additional line ratios involving IR hydrogen recombination lines
(e.g., Pa$\alpha$/H$\alpha$, Br$\gamma$/Pa$\alpha$, and Pa$\alpha$/Pa$\beta$),
can provide better estimates of the internal extinction in highly extinguished
regions as those identified in Arp~299 (see Table \ref{Palfatable}).
 
 The Pa$\alpha$ image of the individual galaxies (see Figs. \ref{ebvic}, \ref{ebvngc}, and AAH00)  has been
 used for measuring all the regions under study with apertures of 1\farcs0 for IC~694 and of 0\farcs9 for
 NGC~3690 to  match our ground-based apertures; the INTEGRAL H$\alpha$ fluxes 
 are given in Tables \ref{tabic} and \ref{tabngc}. In most regions
 extinction estimates based on only optical and optical+infrared lines agree within the uncertainties (Table \ref{Palfatable}). On the other hand, there are two regions, the nucleus of
 IC~694 (A) and C$'$, where the use of the optical+infrared lines indicates a
 larger extinction than using only the optical lines, i.e. A$_{V}$(Pa$\alpha$/H$\alpha$)$\simeq$2$\times$A$_{V}$(H$\alpha$/H$\beta$). This means that the extinctions derived using the H$\alpha$/H$\beta$ ratio are in general consistent with the ones derived using the Pa$\alpha$/H$\alpha$ ratio, except in highly absorbed regions, where there is an underestimate when only the optical lines are used. These
 extinctions are similar with those measured in the M51 nuclear H\,{\sc ii} regions (Calzetti et al. 2005), and in other LIRGs (Alonso-Herrero et al. 2006) using the same near-IR/optical emission lines.

\section{Two-dimensional Ionization Structure}\label{sec6}
\begin{figure*}
\includegraphics[angle=0,width=5.4cm]{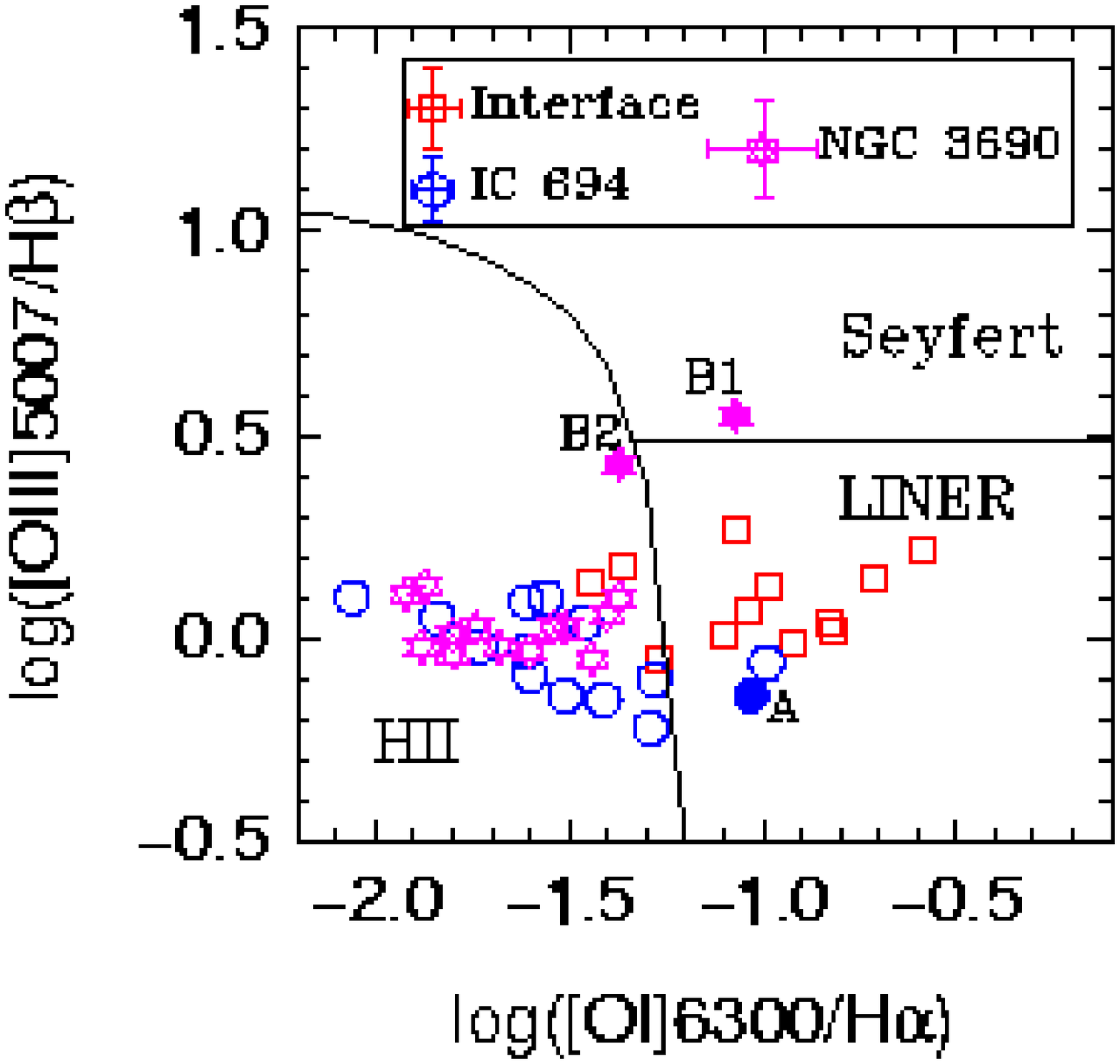}
\includegraphics[angle=0,width=5.4cm]{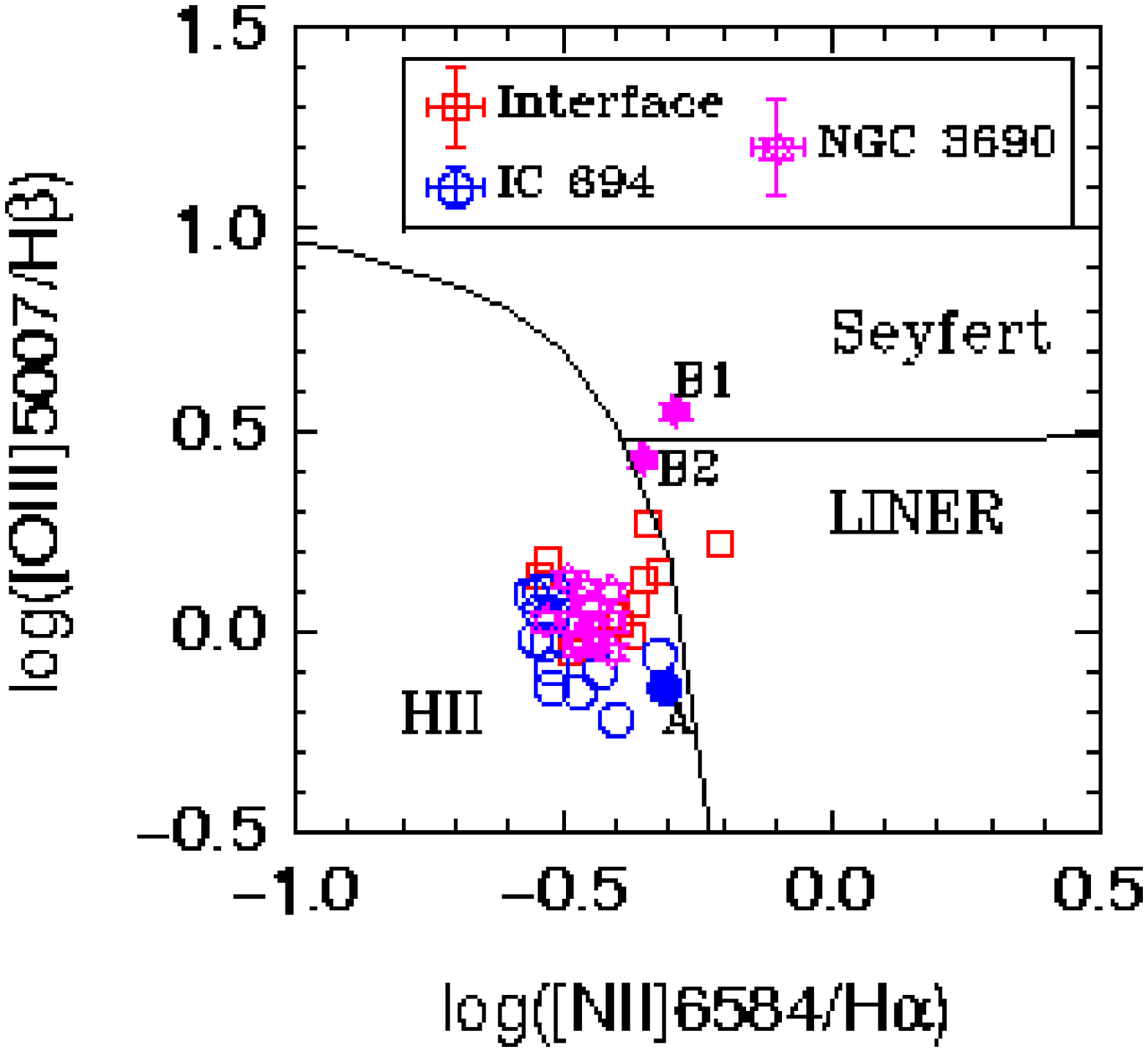}
\includegraphics[angle=0,width=5.4cm]{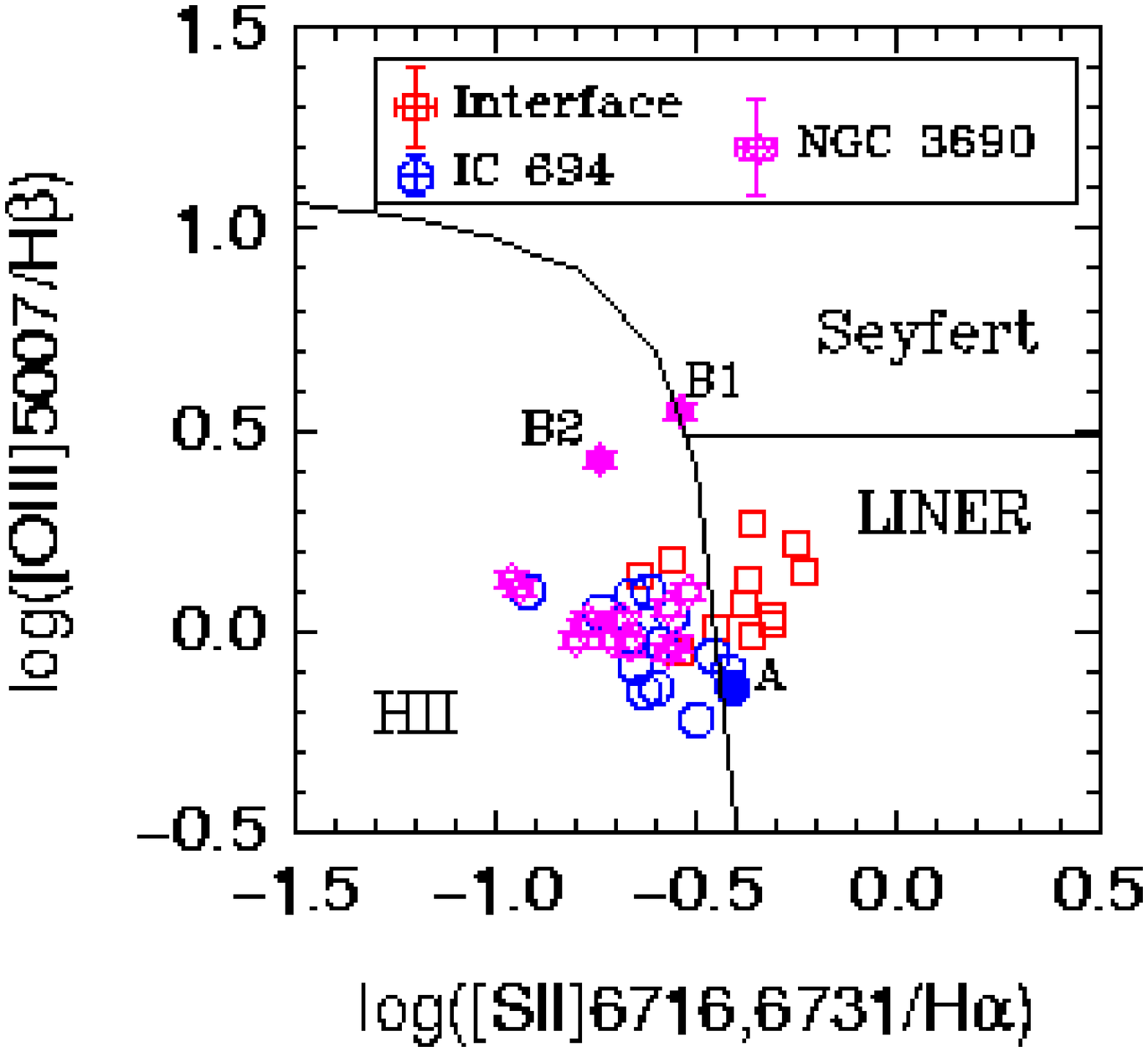}
\caption{Optical emission line diagnostic diagrams for the studied regions of
  Arp~299. Filled symbols represent nuclei A and B1, and region B2. Note that
  B1, as observed with the SB2 bundle, is
classified as Seyfert according to these diagrams. The open symbols
are the regions given in Tables 1, 2, and 3. Typical uncertainties
in the line ratios of each individual galaxy are shown. [\textit{See the electronic edition of the Journal for a color version of this figure.}]}\label{clasicos}
\end{figure*}

The ionization level for the individual regions identified in Arp~299 has been 
derived using standard optical emission line diagnostic diagrams 
(BPT diagrams: Baldwin, Phillips \&
Terlevich 1981; Veilleux \& Osterbrock 1987).  H\,{\sc ii}-like ionization
(i.e. star formation) is present on the majority of the regions of
the individual galaxies, except for nucleus B1 which clearly has Seyfert-like
(i.e. presence of an AGN) state. LINER-like ionization is mainly present on the
interface (Fig. \ref{clasicos}; see also
Tables 1, 2, and 3 for specific values of the emission line ratios). Several regions show
changes in the classification depending on the specific diagram used; these variations have been interpreted as a consequence of changes in the dominant ionizing mechanism from photoionization to shocks
(e.g. Monreal-Ibero et al 2006 and references therein). In particular, the 
[O\,{\sc iii}]$\lambda$5007/H$\beta$-[O\,{\sc i}]$\lambda$6300/H$\alpha$ line ratio is more 
sensitive, and therefore a more reliable tracer, to changes in the ionization conditions  due to fast shocks  
than any of the other optical emission line ratios involving [N\,{\sc
    ii}]$\lambda$6584/H$\alpha$ or [S\,{\sc ii}]$\lambda$6725/H$\alpha$ ratios
(Dopita \& Sutherland 1995). In the following subsections a more
detailed discussion of the two-dimensional ionization structure of the ionized
gas distribution based on the  [O\,{\sc iii}]$\lambda$5007/H$\beta$-[O\,{\sc
    i}]$\lambda$6300/H$\alpha$ diagram and on the [O\,{\sc
    iii}]$\lambda$5007/H$\beta$, and [O\,{\sc i}]$\lambda$6300/H$\alpha$ maps
for the interface, IC~694, and NGC~3690 is given.

\subsection{Interface region: Evidence for an off-nuclear Seyfert-like ionization cone}\label{ionizinterface}

\begin{figure*}
\epsscale{0.35}
\plotone{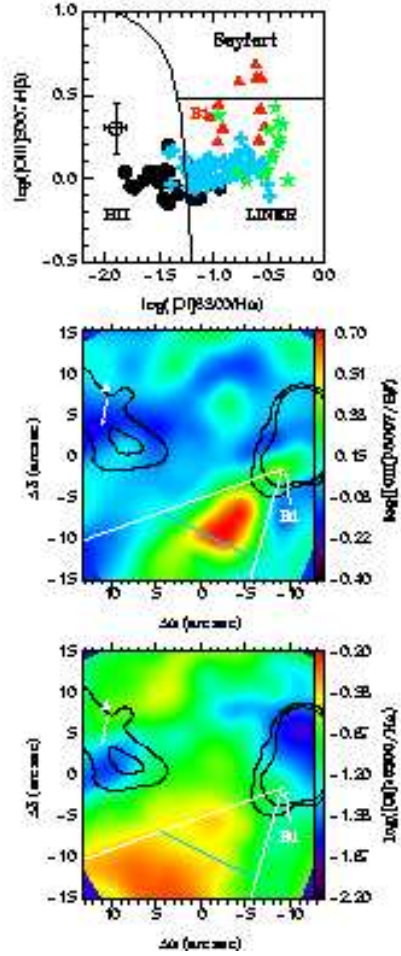}
\caption{Diagnostic diagram and line ratio maps derived from
 INTEGRAL/SB3 data. \textit{Upper panel:} Diagnostic diagram with all the data from the SB3 bundle displayed. The symbol code is as follows: triangles
  represent values within the ionization cone, and at a distance of less
 than 2 kpc from B1. Stars
  represent the zone within the cone at distances of 2 to 4
  kpc from B1. Crosses mark the rest of the regions in the interface. Filled circles represent regions in IC~694 and NGC~3690 (defined as the regions within
 the iso-contours). B1 has been included in the
 ionization cone, and labeled as a triangle. \textit{Center panel:} [O\,{\sc
 iii}]$\lambda$5007/H$\beta$  map.  Nuclei A and B1 are marked.
 The white lines indicate the ionization
 cone. The transversal line to the axis of the cone separates the inner (with Seyfert-like regions) and outer
 (LINER-like) sections of the
 cone. \textit{Lower panel:} [O\,{\sc i}]$\lambda$6300/H$\alpha$
 map. The labels and overlays as in the center panel. [\textit{See the electronic edition of the Journal for a color version of this figure.}]}\label{Diaginterface}
\vspace{0.4cm}
\end{figure*}

The [O\,{\sc iii}]$\lambda$5007/H$\beta$-[O\,{\sc i}]$\lambda$6300/H$\alpha$
 line ratios have been obtained for all the INTEGRAL/SB3 individual spectra
 (Fig. \ref{Diaginterface}). Although these data include the high surface
 brightness nuclear regions of IC~694 and NGC~3690, and their external regions,
 they also cover the area
 that has been identified as the interface zone, i.e. the extended low surface
 brightness region connecting the two galaxies up to distances of about 4 kpc from their nuclei. Changes in the ionization structure traced by the
 [O\,{\sc iii}]$\lambda$5007/H$\beta$ and [O\,{\sc i}]$\lambda$6300/H$\alpha$
 two dimensional spatial distribution
 line ratios are also shown (Fig. \ref{Diaginterface}).   

Three different types of ionization, H\,{\sc ii}-, LINER- and Seyfert-like,
are identified, and the line ratio maps clearly show that the ionization structure has a well defined spatial distribution.
 The majority of the regions associated with the
two individual galaxies (marked with two iso-contours in the figure) are
dominated by H\,{\sc ii}-like ionization, as expected from the intense star
formation taking place there 
(see more details in \S \ref{sec9}). There are some exceptions; the nucleus B1  shows LINER-like activity at this angular resolution. 
 The Seyfert-like ionization is clearly resolved with a size of about 
 7\arcsec\ ($\sim$1.5 kpc) and has a conical morphology, with an opening angle of about 54\,$^{\circ}$ and its apex located in nucleus B1 (at a projected distance of  about 1.5 kpc from the peak of the nebula) within the angular resolution (see Fig.\ref{Diaginterface}).  
Moreover, this highly ionized gas is not associated to any particular concentration of stellar mass according to the optical (WFPC2/F814W; see Fig. \ref{galaxia}) HST continuum image.
 Off-nuclear Seyfert-like nebulae at distances of  few kpc from the nucleus have already been reported in other (U)LIRGs with Seyfert nucleus (Mrk~273: Colina et al. 1999), and interpreted as photoionization of extranuclear  interstellar gas by the AGN. Therefore, our data  strongly suggest photoionization by a radiation
  cone escaping from a central dust-enshrouded AGN source located in B1.
The highly ionized conical
    structure in Arp~299 (NGC~3690) is also detected at distances of up to 4 kpc from B1; however, the ionization 
associated to these outer regions of the cone is LINER-like rather than Seyfert (Fig. \ref{Diaginterface}). Therefore, it seems that within the cone there are two ionization regimes well separated spatially. At 
projected distances from B1 smaller than about 2 kpc the ionization is likely due to radiation coming from the AGN located in B1, while at larger distances LINER-like ionization is dominant. This LINER-like  ionization is also found in other areas of the interface region, however with a lower degree of excitation, as traced by the [O\,{\sc
    i}]$\lambda$6300/H$\alpha$ ratio measured in these regions.

\subsection{IC~694}\label{ionizic}
\begin{figure*}
\epsscale{0.5}
\plotone{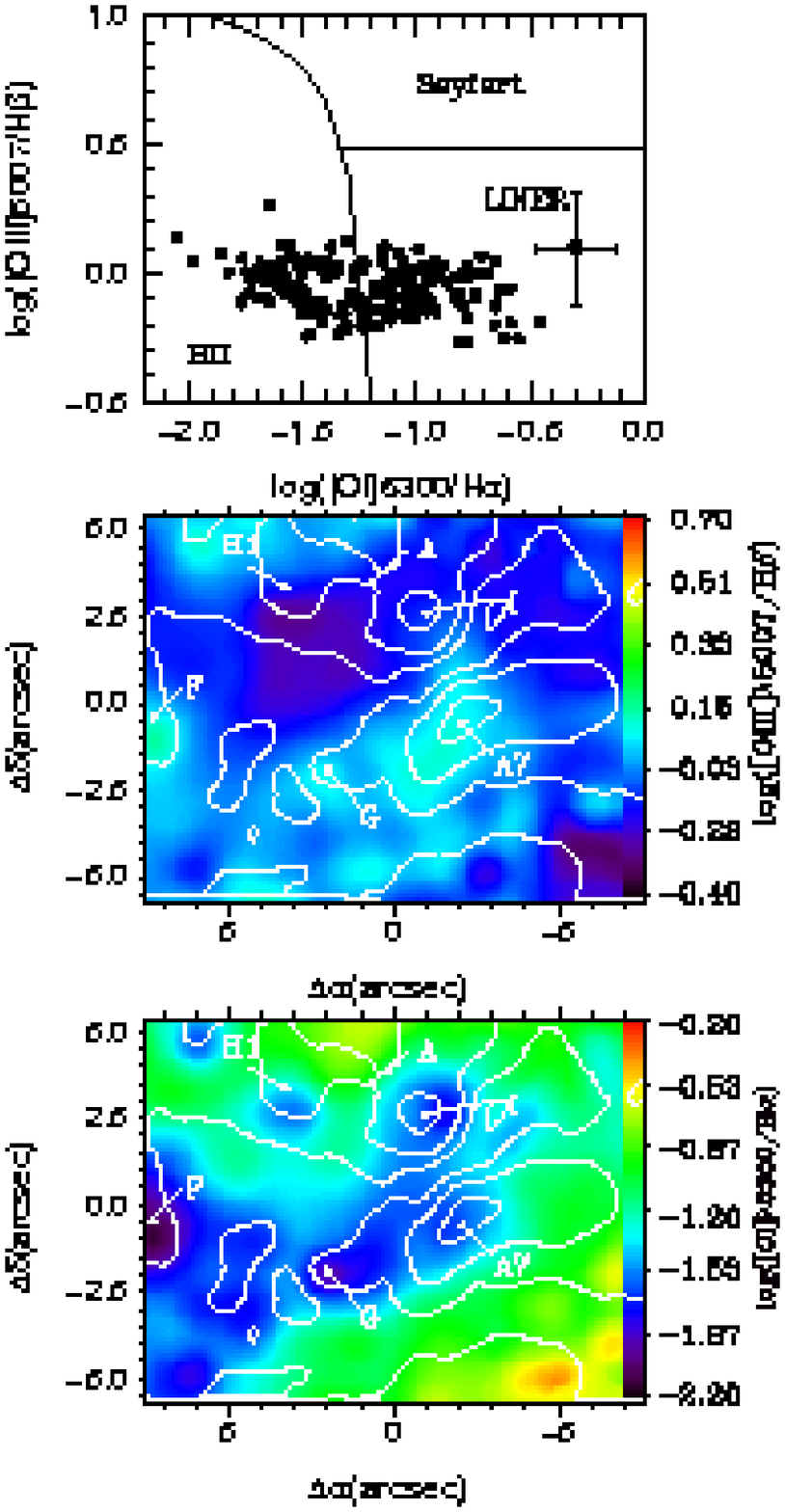}
\caption{Emission line diagnostic diagram and excitation maps based on
 line ratios for IC~694. \textit{Upper panel:} Diagnostic diagram
 obtained for the individual SB2 spectra. \textit{Center and Lower
 panels:} Diagnostic maps derived from reconstructed emission line maps. As a reference we have marked several
 regions and superimposed red continuum contours. [\textit{See the electronic edition of the Journal for a color version of this figure.}]}\label{diagic}
\vspace{0.4cm}
\end{figure*}

Over the area covered by the INTEGRAL SB2 bundle, H\,{\sc ii} and LINER-like
ionization are present in this galaxy (Fig. \ref{diagic}).
The main body of the galaxy, including the regions
close to the nucleus A and the spiral arm structure (A7, G) towards
the east (F), is dominated by the star-forming knots. LINER-like ionization is
present elsewhere, being more prominent  in diffuse low surface
brightness regions located at a projected distance of about 1.8 kpc. The [N\,{\sc ii}]$\lambda$6584/H$\alpha$  and  the [S\,{\sc ii}]$\lambda$6725/H$\alpha$
line ratio maps (not shown) also have a similar structure though, as explained
above, they tend to indicate a lower ionization.  
  
\subsection{NGC~3690}\label{ionizngc}
\begin{figure*}
\epsscale{0.35}
\plotone{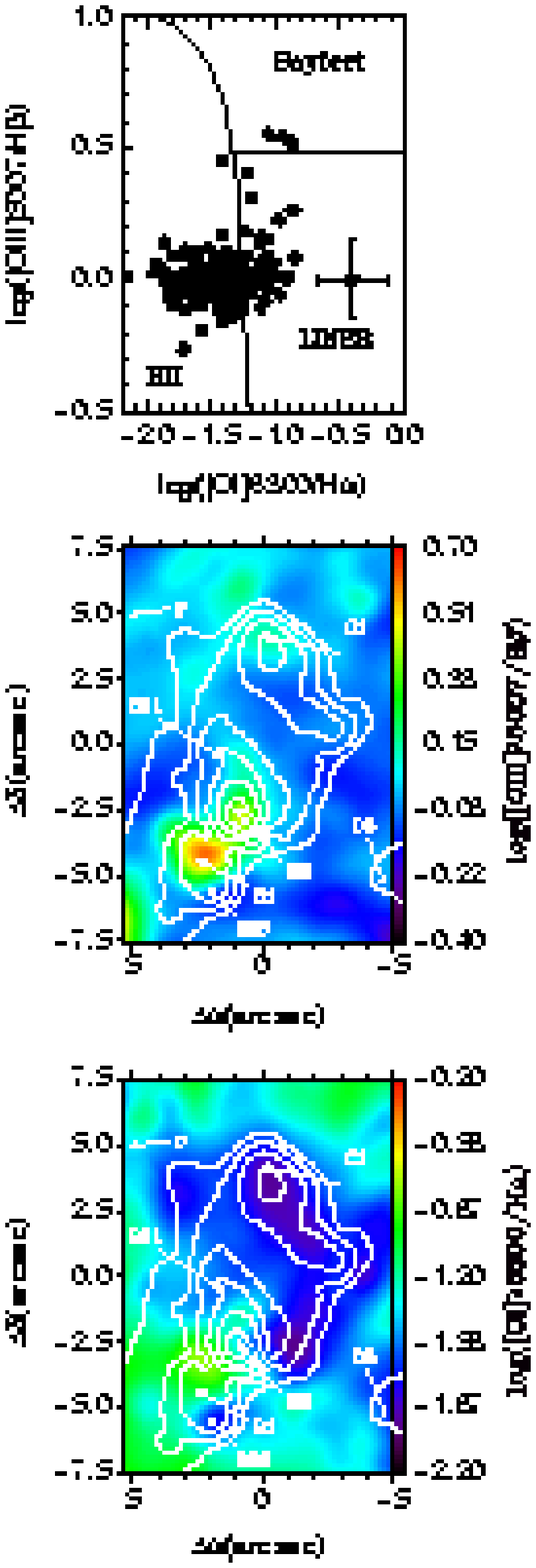}
\caption{Emission line diagnostic diagram and excitation maps based on
 line ratios for NGC~3690. \textit{Upper panel:} Diagnostic diagram
 obtained from the individual SB2 spectra. \textit{Center and Lower
 panels:} Diagnostic maps derived from reconstructed emission line maps. As a reference we have marked several
 regions and superimposed red continuum contours. [\textit{See the electronic edition of the Journal for a color version of this figure.}]}\label{diagngc}
\end{figure*}

Three types of excitation level (H\,{\sc ii}-, LINER- and Seyfert-like) are present in this individual galaxy, as shown in the
[O\,{\sc iii}]$\lambda$5007/H$\beta$-[O\,{\sc i}]$\lambda$6300/H$\alpha$
diagnostic diagram and in the two dimensional maps
(Fig. \ref{diagngc}). Regarding the nucleus B1, these higher resolution data
give a clear Seyfert-like classification in all these line ratios  (Fig. \ref{clasicos}), and show
that there is an extended zone of about 2\arcsec~(0.4 kpc) in length
associated to B1 with Seyfert characteristic. This is the inner section of the
AGN-like ionization cone
detected with the SB3 bundle, with its apex coincident with B1. Moreover, the region B2 presents also a high ionization state in the border line of Seyfert/LINER, characteristic that could indicate the presence of the counter ionization cone. 
The detection of Seyfert-like activity based on optical emission lines
represent the first evidence in the optical for the presence of an AGN in B1, and
supports the recent claim of an AGN based on X-ray (Della Ceca et al. 2002) and
mid-IR (Gallais et al. 2004) data. Previous spectroscopic studies in the
optical (Gehrz et al. 1983; Coziol et al. 1998; Keel 1984) did not find AGN
emission, probably due to the large size of the slits (3\farcs6,
2\farcs5, and 8\farcs0). This is in agreement with our results, where the Seyfert nature of the B1 nucleus has been diluted in the SB3 (\o = 2\farcs7) spectra, whereas in the higher spatial resolution SB2 (\o = 0\farcs9) spectra the classification is clear. 
 
H\,{\sc ii}-like ionization dominates the rest of the main body of the
galaxy (e.g. complex C), while LINER-like is present in a few outer regions.
The [O\,{\sc i}]$\lambda$6300/H$\alpha$ map has its peak displaced about
0\farcs7 (0.14 kpc) to the north with respect to B1. Similar displacements
are observed in the [N\,{\sc ii}]$\lambda$6584/H$\alpha$ and in the [S\,{\sc
    ii}]$\lambda$6725/H$\alpha$ maps. These lines and in particular [O\,{\sc
    i}]$\lambda$6300/H$\alpha$ are a good tracers of shocks, thus these
shifts may indicate different ionization mechanisms around B1 on scales of 200 pc. These small scale ionization changes need further investigation with high angular resolution spectroscopy.

\section{Electron density}\label{electron}
The two-dimensional electron density maps of the ionized gas distribution have
been derived for the entire Arp~299 system using the standard  [S\,{\sc{ii}}]$\lambda$6716/[S\,{\sc{ii}}]$\lambda$6731 lines ratio (Aller 1984; Osterbrock 1989). 
The average line ratio for the interface and the individual galaxies show a similar value, about 1.3$\pm$0.1, indicating an average uniform electron density ($\sim$100 cm$^{-3}$) over the entire Arp~299 system. However, the ionization cone detected in NGC~3690 presents a well defined gradient of this line ratio value, that roughly ranges from 1.2$\pm$0.15 in the higher excitation (i.e. Seyfert-like) center of the cone, to 1.8$\pm$0.15 in the lower excitation (i.e. LINER-like) outer section. Thus, the electron density seems to decrease from about 180 cm$^{-3}$ at a distance of 1.5 kpc from B1 to less than 10 cm$^{-3}$ at about 4 kpc.

\section{The origin of the LINER-like ionization}\label{sec8} 
\begin{figure*}
\includegraphics[angle=0,width=5.4cm]{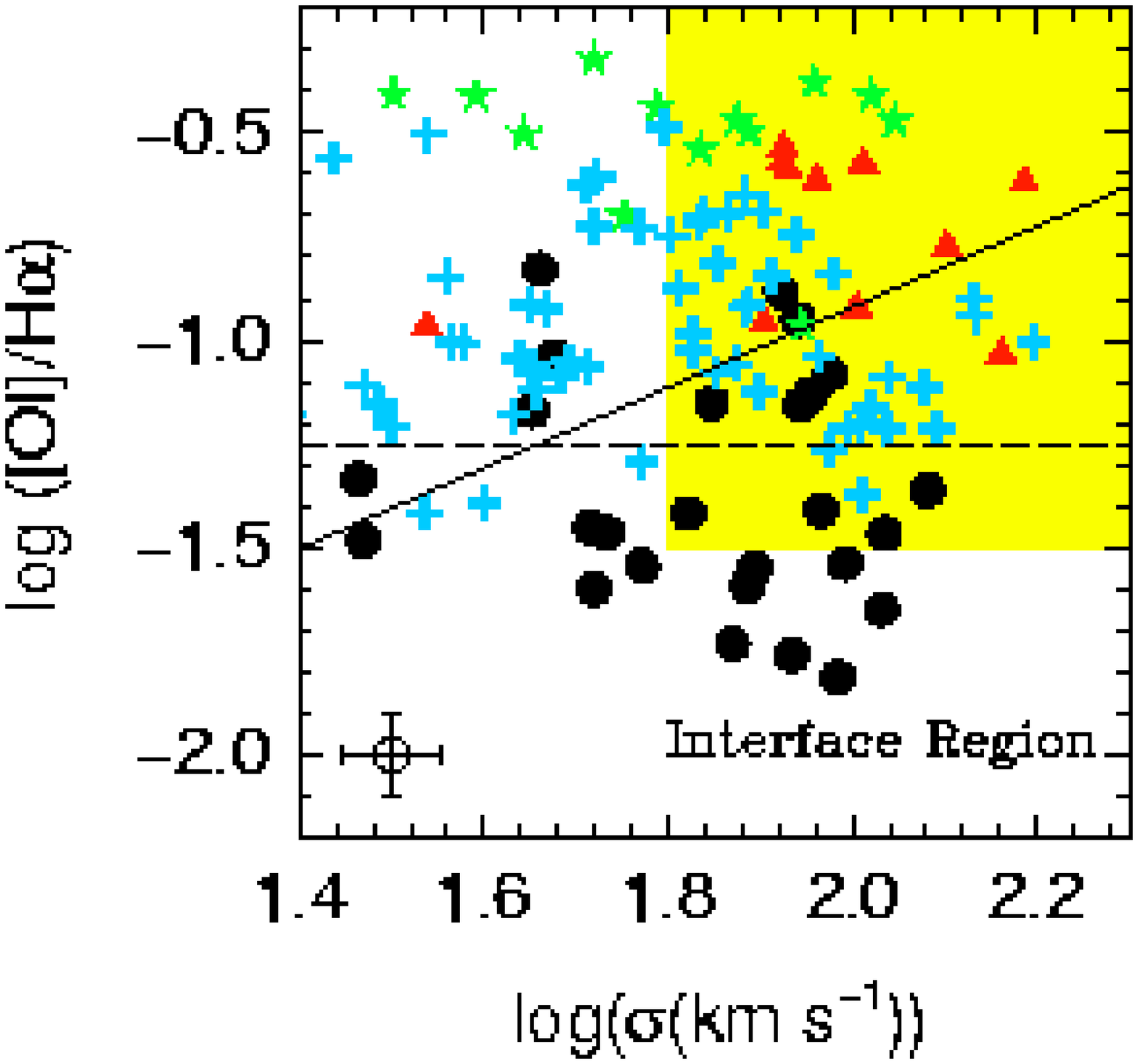}
\includegraphics[angle=0,width=5.4cm]{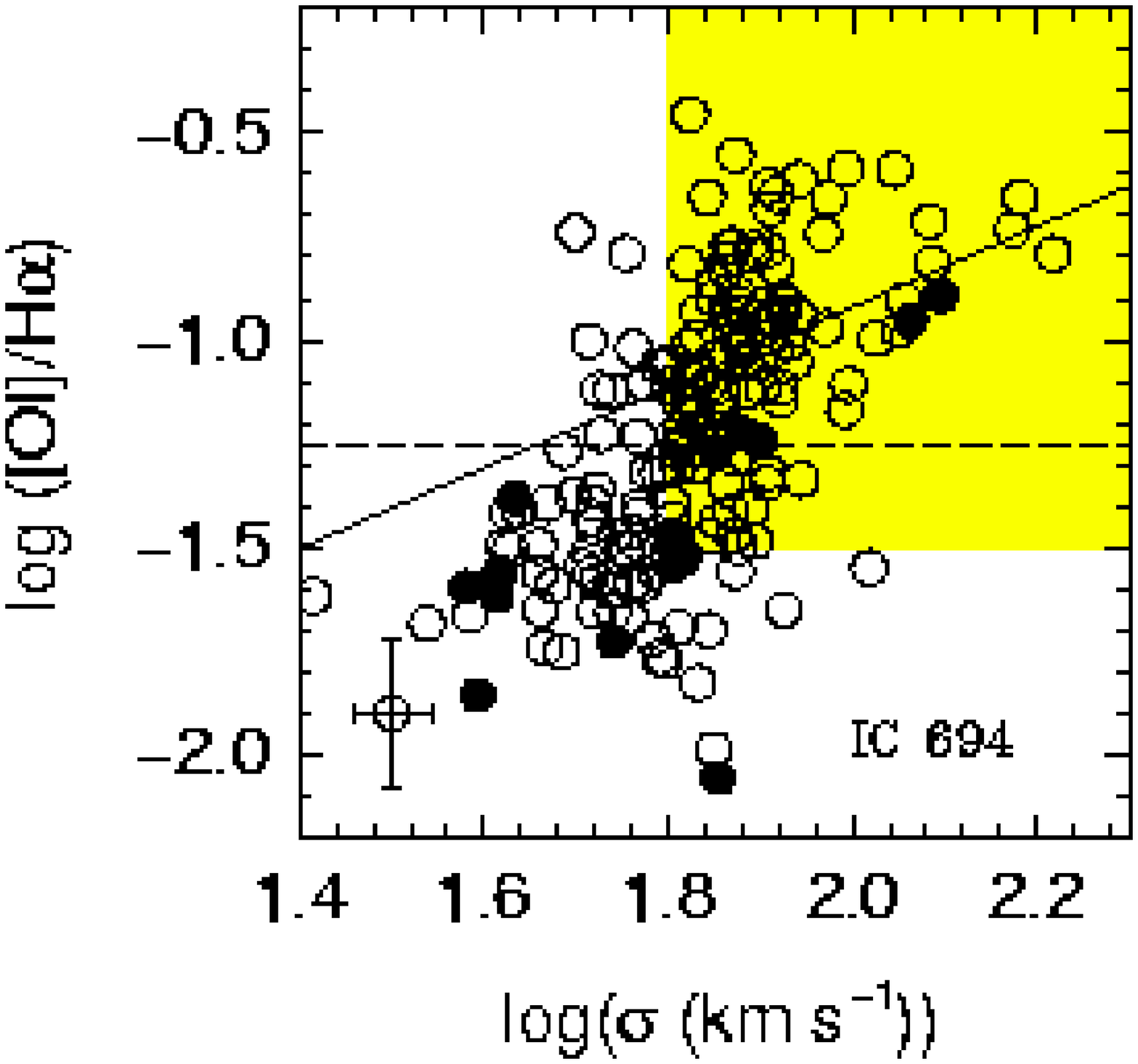}
\includegraphics[angle=0,width=5.4cm]{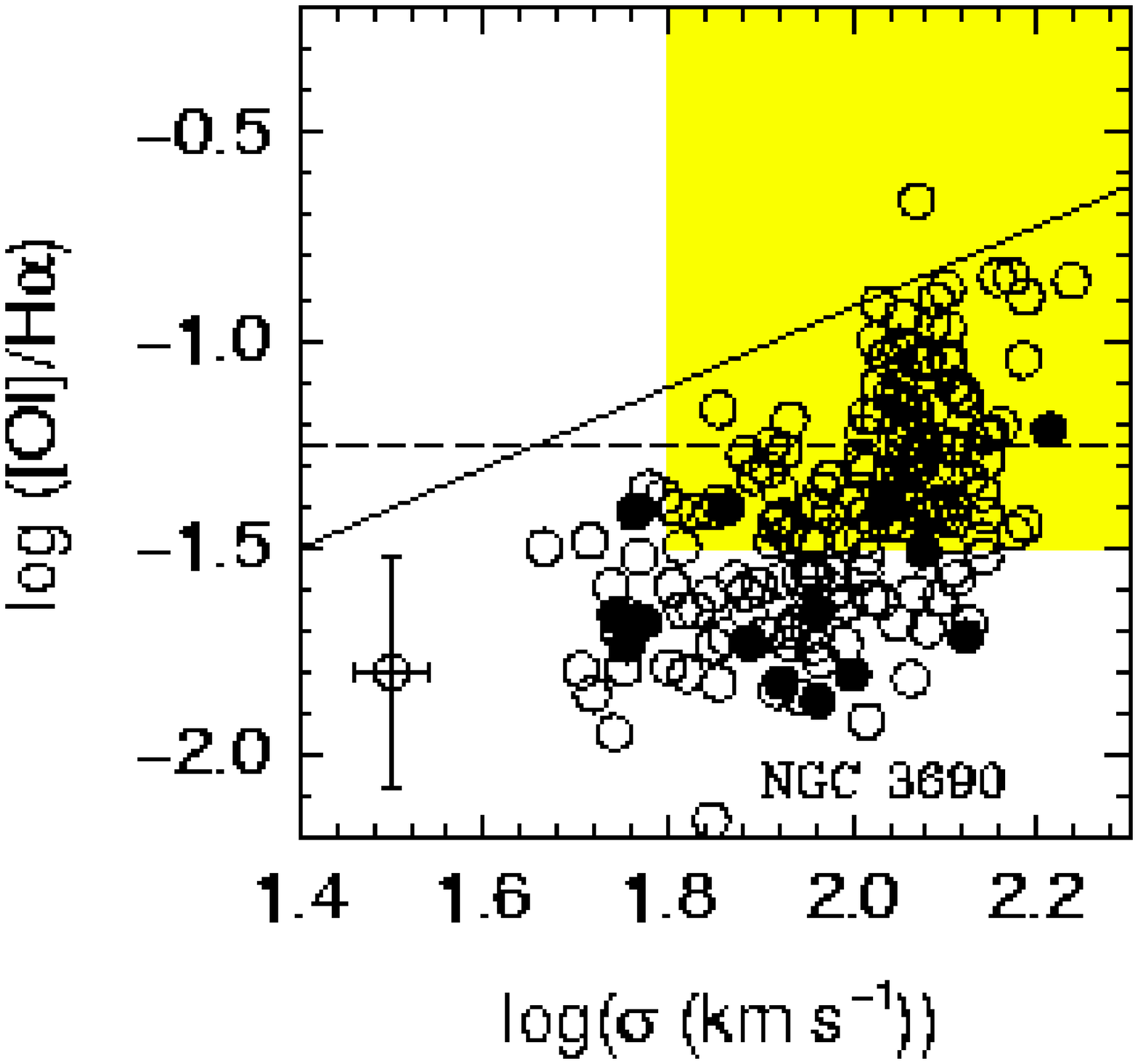}
\caption{Relation between velocity dispersion and
  [O\,{\sc{i}}]$\lambda$6300/H$\alpha$ ratio for the three INTEGRAL
  pointings, i.e. interface region (\textit{Left}), IC~694 (\textit{Center panel}) and
  NGC~3690 (\textit{Right panel}). In all cases the relation derived by Monreal-Ibero
  et al. (2006) for a set of ULIRGs is shown (solid line). The shaded area indicate the region covered by extra-nuclear regions of ULIRGs (Monreal-Ibero et al. 2006). The
  horizontal dashed line marks the average values separating H\,{\sc ii} and
  LINER-like ionization. \textit{Left panel:} The symbol code as in
  Fig. 10. \textit{Center and Right panels:} Filled symbols indicate
  the fibers of the studied individual regions listed in Tables \ref{tabic} and \ref{tabngc}. [\textit{See the electronic edition of the Journal for a color version of this figure.}]}\label{correlacion}
\end{figure*}

Most regions of the Arp~299 system have been classified as LINER-like, although the mechanism producing this ionization needs to be determined. Some authors have found in (U)LIRGs a positive correlation between the [O\,{\sc i}]$\lambda$6300/H$\alpha$ ratio and the velocity dispersion (Monreal-Ibero et al. 2006), which has been considered as evidence supporting the shocks as the origin for the ionization (Armus et al. 1989; Dopita \& Sutherland 1995).
The [O\,{\sc i}]$\lambda$6300/H$\alpha$ versus the H$\alpha$ velocity
 dispersion for all the spectra obtained with INTEGRAL has been represented (Fig. \ref{correlacion}) in order to investigate the origin of the LINER-like ionization in Arp~299. The four main regions of the system observed with the INTEGRAL SB2 and SB3 bundles are analyzed separately: the inner and outer sections of the ionization cone, the interface region and finally the individual galaxies. The outer section of the ionization cone (stars in Fig. \ref{correlacion}, \textit{Left panel}. LINER-like excitation , i.e.[O\,{\sc i}]$\lambda$6300/H$\alpha$ $>$ -1.25) is independent of the velocity dispersion and present a stable ionization level, while with a wider variety in the ionization level, the inner section of the cone (Seyfert-like, triangles in Fig. \ref{correlacion}, \textit{left panel}) does not follow the correlation founded in external regions of ULIRGs (Monreal-Ibero et al. 2006). The interface region (crosses in Fig. \ref{correlacion}, \textit{Left panel}) present a stable LINER-like excitation level, and is not dependent of the velocity dispersion.

 With regards to the individual galaxies IC~694 and NGC~3690 as observed with the lower angular resolution (SB3) bundle (filled circles in Fig. \ref{correlacion}, \textit{Left panel}), the ionization is mainly H\,{\sc{ii}}-like, with no evidence of a kinematical correlation. For the higher angular resolution (SB2) data (see Fig. \ref{correlacion}, \textit{Center and Right panels}), the actual measurements of the individual galaxies do not follow the correlation, although they present a well defined clustering.  The galaxy IC~694 has a mean log$\sigma$$\sim$1.8, and presents about the same fraction of H\,{\sc{ii}} and LINER activity. In the case of NGC~3690 the velocity dispersion is clustered around log$\sigma$$\sim$2, and there is a large fraction of regions with H\,{\sc{ii}}-like ionization.

Clearly the present data show a different behavior than that found in ULIRGs
(Monreal-Ibero et al. 2006), and this is probably due to the fact that on each
case the measurements are done over regions located at different distances to
the nucleus, and with different characteristics. The original correlation was
derived for diffuse extra-nuclear regions in ULIRGs, which are located at
distances of about 5$-$15 kpc from the nucleus of the galaxy. In the present
case the interface region and the ionization cone have regions at distances of
up to 4 kpc from the nucleus, and comparing with the original limits of the
correlation  (see shaded area on Fig. \ref{correlacion}, \textit{Left panel})
the data of the interface and both regions of the ionization cone lie in
general above the ratio, i.e. this regions show higher excitation for the same velocity dispersion.
In the case of the individual galaxies observed with the actual higher resolution bundle (Fig. \ref{correlacion}, \textit{Center and Right panels}) the present data are closer to the nucleus than in the case of ULIRGs (up to $\sim$2.5 kpc) and in most cases the regions involved are compact bright sources rather than diffuse. Comparing with the original limits of the relation (see shaded area on Fig. \ref{correlacion}, \textit{Center and Right panels}) the IC~694 data are clustered around, while the ones of NGC~3690 are below the correlation. This means that in the present case for the individual galaxies the star formation influence is still present, and that the velocity dispersion close to the nucleus traces mass rather than flows/shocks.

Given an ionization mechanism, the ionization state of the surrounding gas can
be changed as a result of the variations in the electron density, that affect to the ionization parameter (\emph{U}$\propto$N$_{e}$$^{-1}$r$^{-2}$). This appears to be the case in the outer high excitation cone associated to NGC 3690. As already mentioned, the electron density seems to have a radial variation within the cone, and by having the exact variation of N$_{e}$ with r, the radial dependence of the parameter \emph{U} would be described more precisely. Unfortunately in the present case the exact radial dependence of N$_{e}$ cannot be derived; in spite of that, the variation detected on the electron density, and therefore the LINER activity could be accounted for by the influence of an AGN (if N$_{e}$ $\propto$ r$^{-\alpha}$). The variation of N$_{e}$ implies a variation on the ionization parameter (\emph{U} $\propto$ r$^{-2+\alpha}$), and thus the low excitation is valid if the source is not local.

 In conclusion, for the Arp~299 system studied here there are no clear evidences for tidally induced shocks as found in ULIRGs. Therefore, the origin for the LINER ionization is unclear, and at least in some zones (i.e. outer parts of the cone) it could, perhaps, be due to a radiation
field similar to that of a low-luminosity AGN.

\section{Star formation properties}\label{sec9}
It is well known that merging processes between gas rich galaxies can trigger star formation. As predicted by numerical simulations (Mihos \& Hernquist 1994), the process can vary depending on whether or not the parent disk galaxies have a bulge. The star-forming process at a low level (SFR$\sim$20 M$_{\odot}$ yr$^{-1}$) consequence of the merging of two disk galaxies without a bulge is produced early on in the merging phase, and can be sustained for about 150 Myr. On the other hand, in the merger of galaxies with a bulge the starburst (SFR $\sim$ 50$-$100 M$_{\odot}$ yr$^{-1}$) is produced later in the merging phase, and lasts for  about 50 Myr.

To investigate the properties of the young stellar populations of a galaxy assuming instantaneous and continuous bursts, synthetic models such as SB99
(Leitherer et al. 1999; Vazquez \& Leitherer 2005) can be used. In the following the stellar population synthesis model considers a Salpeter initial mass function (IMF), with lower and upper mass limits of 0.1~M$_{\odot}$ and 100~M$_{\odot}$, the Geneva tracks with high mass loss, and the Calzetti extinction law (Calzetti et al. 2000). Instantaneous bursts are normalized to 10$^{6}$~M$_{\odot}$, while the continuous star formation is normalized to a star formation rate of 1 M$_{\odot}$ yr$^{-1}$. Combining the information from the two dimensional distribution of the H$\alpha$ equivalent width and luminosity, and the
photometric results, parameters such as the extinction to the gas, the age of the burst,  and the stellar mass for each individual region can be derived. For the present analysis we use the INTEGRAL SB2 bundle with a resolution of 0\farcs9 (i.e. 0.18 kpc), thus in the majority of cases the data include both the contribution from the stellar clusters and from a diffuse component.
 
\subsection{Photometry of the regions of IC~694 and NGC~3690}
The aperture photometry as derived by AAH00, using the HST NIC1
F110M, NIC2 F160W, NIC2 F222M images is used in the following analysis. In addition optical (WFPC2/F814W) photometry over the same regions, using the same
apertures, and after performing the background subtraction and the aperture
correction (for the latter see Holtzman et al. 1995) has been obtained. The apertures used were 0\farcs9 diameter
for the nuclei of IC~694 and NGC~3690 A and B1, and for the star-forming region B2,
and 0\farcs5 diameter for the rest of the star-forming regions. The HST magnitudes and colors derived are shown in Table
\ref{Magtable}. In addition, the HST magnitudes and colors were modeled by convolving the simulated galaxy spectral energy distribution with the system (optics, filter and detector) transmission function.

\subsection{Properties of the H$\alpha$ emitting regions}
 In the following, instantaneous bursts models characterized by the predicted H$\alpha$ luminosity and equivalent width, and near-IR colors will be used
to study the stellar populations of the H$\alpha$ emitting regions under analysis. Comparing the measured values with the models, it is possible to derive the age, extinction and mass of the individual regions under study. 
Regarding to the  nucleus B1, and as previously indicated, it is classified as Seyfert, and the fraction of the ionizing flux coming from the buried AGN in this region is not well determined. Therefore the conclusions regarding the star formation properties in B1 based on the H$\alpha$ luminosity have to be taken with caution, and represent an upper limit. Because of this fact, the nucleus B1 is   basically not included in the following discussion, although the properties derived are listed in the tables.

\begin{figure*}
\includegraphics[angle=0,width=7.4cm]{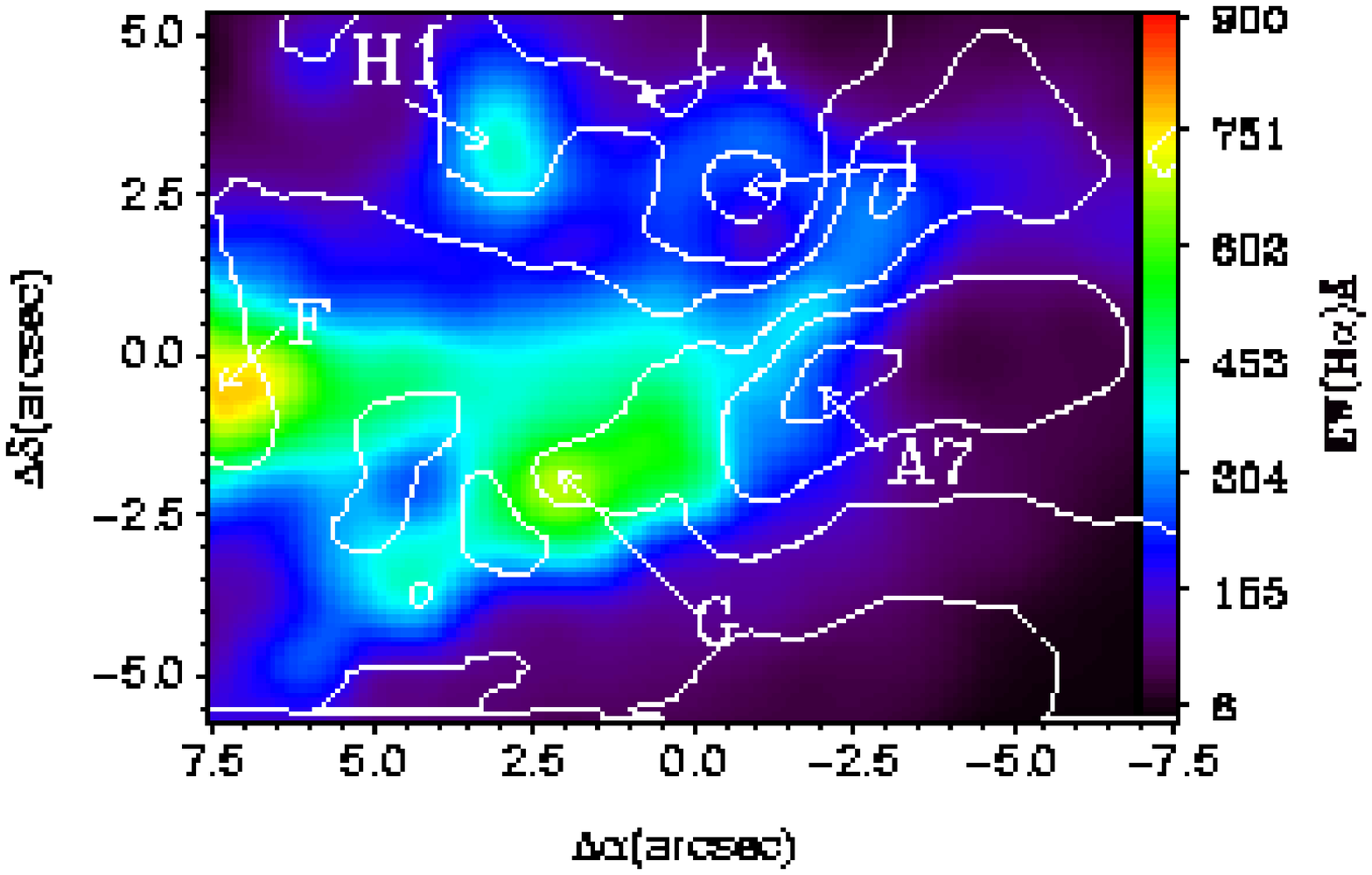}
\includegraphics[angle=0,width=7.0cm]{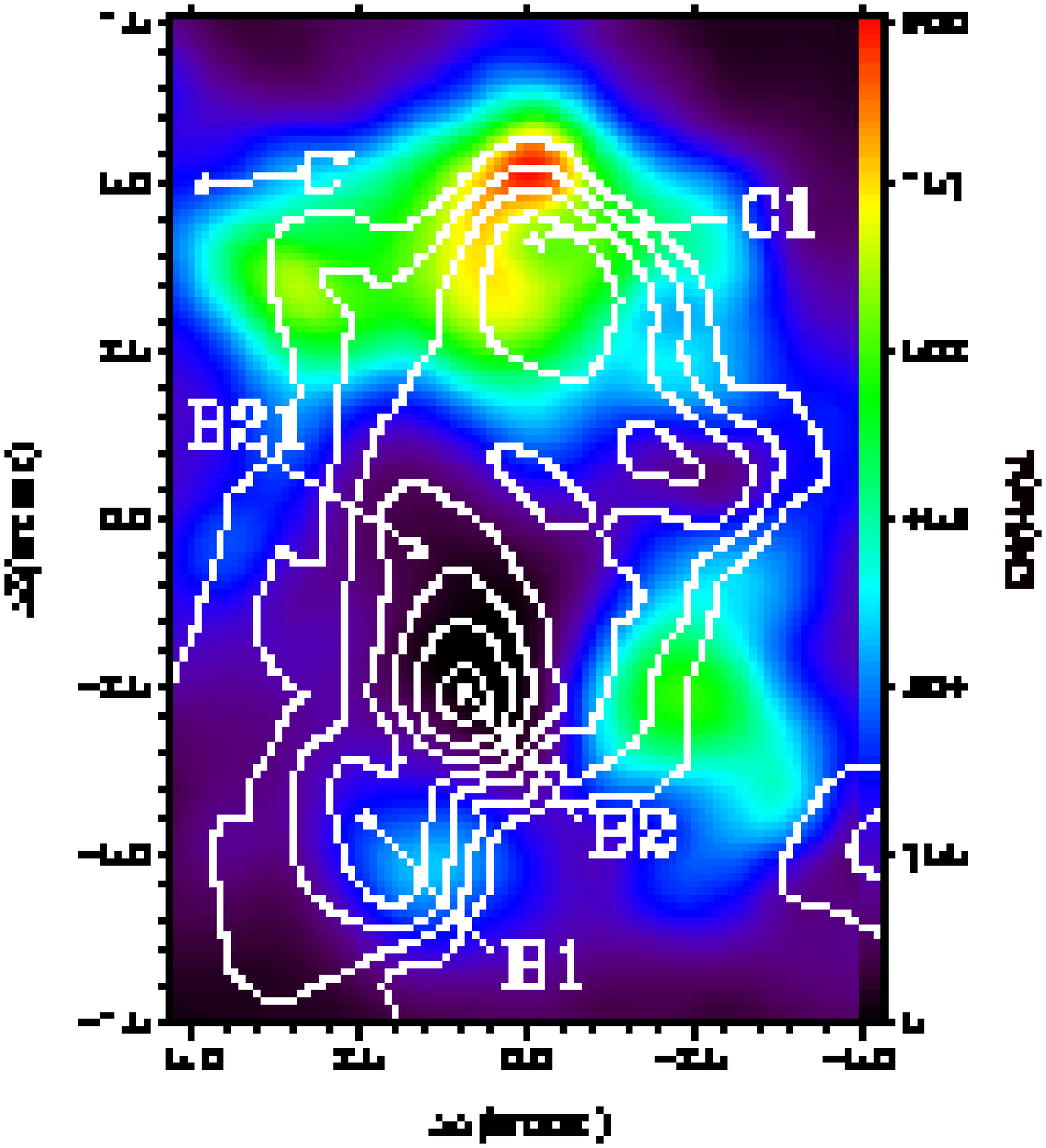}
\caption{H$\alpha$ equivalent width maps. Orientation is north up,
  east to the left. \textit{Left panel:} IC~694. \textit{Right panel:} NGC~3690. Note that nuclei A and B1 have low equivalent widths. [\textit{See the electronic edition of the Journal for a color version of this figure.}]}\label{ew}
\vspace{0.4cm}
\end{figure*}

The H$\alpha$ luminosity in specific regions with large H$\alpha$ equivalent width (Fig. \ref{ew}) traces preferentially the presence of young stellar populations ($<$5$-$10 Myr). However, the presence of an old underlying population tends to decrease the
EW, and therefore to underestimate the age of the young stellar population. In the case of IC~694 and NGC~3690 the respective nuclei A and B1, and the NGC~3690 brightest optical region B2 present low EW(H$\alpha$).
The ages derived from the EW(H$\alpha$) for several individual regions are shown in Table \ref{tablamodic}.

 With an age for the nucleus (A) of 5.3$\times$10$^{6}$ years, IC~694 does not show a wide spread in age for the individual star-forming regions; almost all are around 5$\times$10$^{6}$ years, except for I1 with 5.9$\times$10$^{6}$ years and the youngest regions, concentrated towards the east of the spiral-like arm (F
and G, with 3.3 and 3.4$\times$10$^{6}$ years respectively). The galaxy NGC~3690 shows an average age for the individual regions of $\sim$5.5$\times$10$^{6}$ years, with the eldest one (7.2$\times$10$^{6}$ years) located towards the north of B2. The youngest sources are located in the northern part of the galaxy, and correspond to complex C. Independent estimates (Satyapal et al. 1999) also based on instantaneous bursts and characterized by a Salpeter IMF, derived ages of about 6$\times$10$^{6}$ years for the main components of the Arp~299 system, that are compatible with the results obtained here. Compared to AAH00, who used a different model with Gaussian bursts, the present age for B2 is younger, but applying their EW(Br$\gamma$)$\leq$5 \AA\ to our model, the age for B2 is $\geq$7$\times$10$^{6}$ years, compatible with the actual calculations.

\begin{figure*}
\epsscale{1.0}
\plottwo{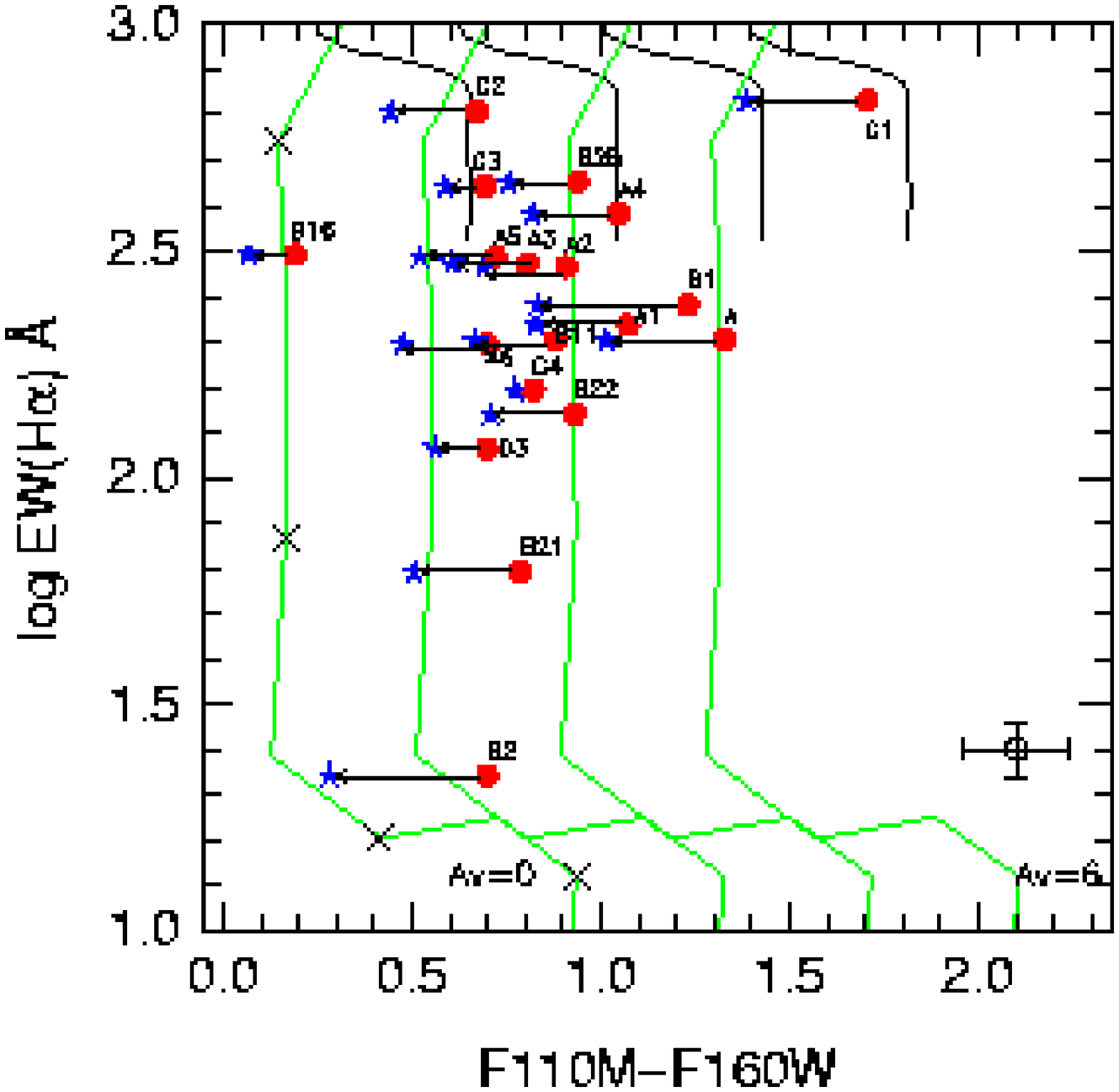}{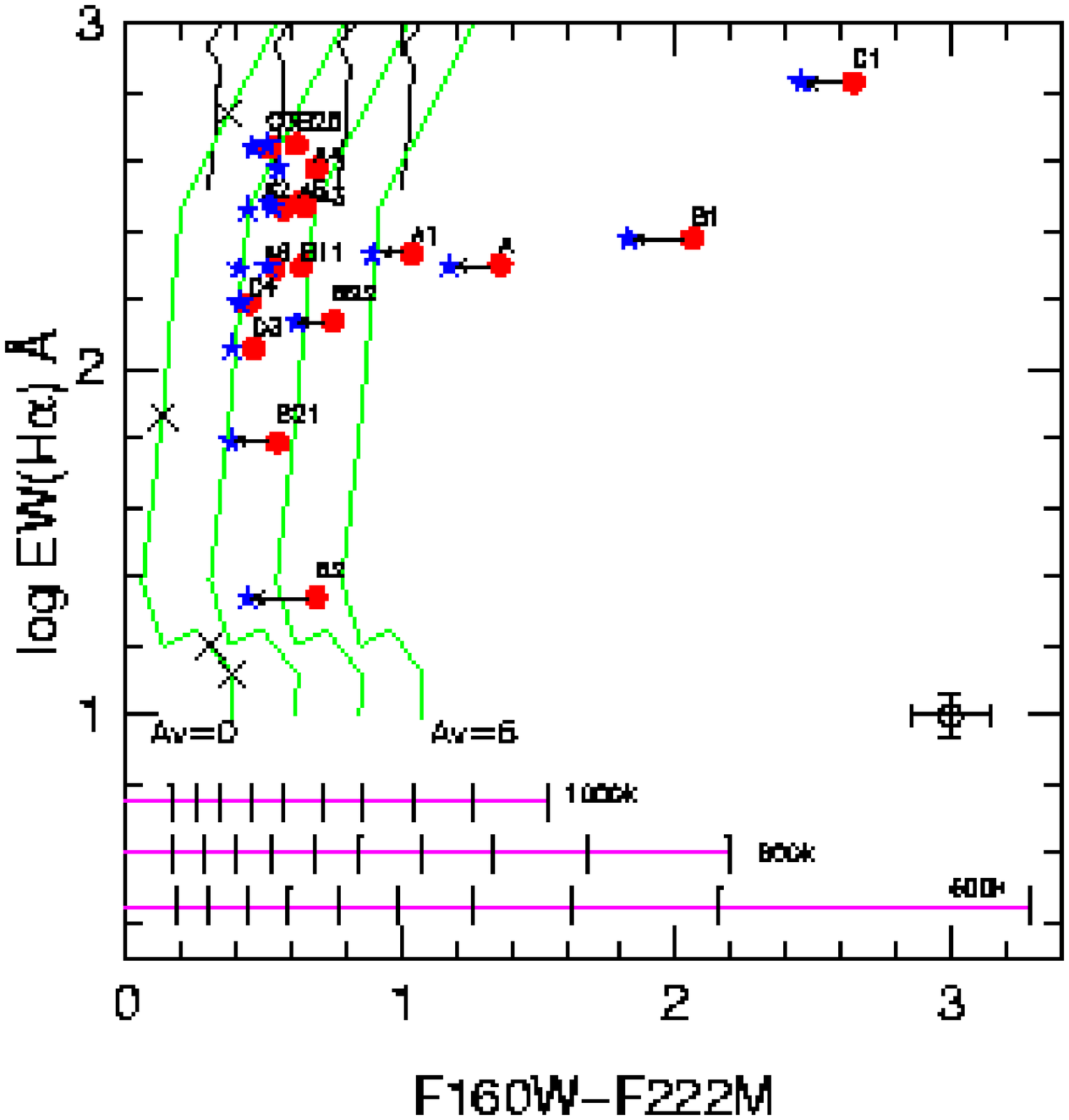}
\caption{Plots representing the results of the stellar evolutionary
  models (SB99, Geneva High Isochrones, Salpeter IMF, using upper and lower mass boundaries of 0.1 and 100 M$_{\odot}$) compared with the measured
  values. The long tracks represent the evolution of instantaneous bursts
  as a function of age (from top to bottom, black crosses mark 4, 6, 8 and 10 Myr). Short tracks correspond to continuous bursts using a SFR of 1
  M$_{\odot}$ year$^ {-1}$. In both cases from left to
  right the extinction ranges from A$_{V}$=0 to A$_{V}$=6, in steps of 2 magnitudes. The filled circles are the
  measured values, and the stars represent the extinction corrected values from the H$\alpha$/H$\beta$ ratio. \textit{Left panel:} F110M-F160W color vs H$\alpha$ equivalent
  width. \textit{Right panel:} F160W-F222M color vs H$\alpha$ equivalent
  width.  The long horizontal lines represent the colors expected from a extinction-free stellar population of 6$\times$10$^{6}$ years with an increasing K-band flux contribution of hot dust at different temperatures (from 0\% to 100\%, increasing 10\% on each step). In both cases nucleus B1 is included for completness. [\textit{See the electronic edition of the Journal for a color version of this figure.}]}\label{Genevamodels1}
\end{figure*}

Extinction and dust properties of the star-forming regions can be studied using the near-IR colors F110M-F160W (approximately \textit{J-H}) and F160W-F222M (approximately \textit{H-K}) together with the EW(H$\alpha$). Under the assumption that the extinction to the stars is equivalent to the extinction towards the gas, the EW(H$\alpha$) value is independent of A$_{V}$ (Fig \ref{Genevamodels1}). We have unreddened both observed colors using a foreground screen model and the extinction derived from the H$\alpha$/H$\beta$ ratio, and compared with the models. Once corrected for extinction, and taking into account the error bars, in the majority of the regions the values are close to the predicted A$_{V}$=2$-$3 mag track. Thus in general the extinction derived from the H$\alpha$/H$\beta$ ratio presented in Tables \ref{tabic} and \ref{tabngc} show an offset with respect to the ones predicted by the colors. There are two possible explanations for this discrepancy; on the one hand the extinction derived from the H$\alpha$/H$\beta$ ratio can be underestimated (see \S \ref{sec54}), on the other hand the discrepancy traces the contribution of hot dust emission. For the highly reddened regions (e.g. A, B1\footnote{In this discussion related to the hot dust the nucleus B1 is included, because even with high errors in the age derived from the EW(H$\alpha$), the color excess exists and needs to be explained.}, C1) the colors measured (and corrected by extinction) are very red. As already mentioned (and discussed by AAH00) one of the most plausible explanations for this color excess is the presence of hot dust (T$>$600 K), which would mainly affect the F160W-F222M color. In an attempt to take into account the contribution of the hot dust emission, the $\lambda^{-2}$$\cdot$B$_{\lambda}$
wavelength modified black-body thermal emission (Hildebrand 1983) has been used. The results demonstrate that depending on the temperature (which has not necessary to be homogeneous over the line of sight) and the amount of dust, the contribution of hot dust ($\sim$800 K) emission could represent, on average, about 80\% of the K-band emission (Fig. \ref{Genevamodels1}, \textit{Right panel}) for the highly extincted regions (A, B1, C1), and about 30\% for the rest of the regions. 
 This result supports our previous conclusion that the dust is more concentrated towards the nuclei and some particular regions, than in the extra-nuclear star forming regions.

An alternative to the presence of a hot dust emission is that the structure of
the dust-gas distribution does not follow the simple screen geometry, but it
is closer to a clumpy foreground or a mixed dust-star distribution. Assuming a
model where the gas is homogeneously mixed with the stars, the effect of an
A$_{V}$=6 mag over the F160W-F222M and F110M-F160W measured colors will
produce bluer colors: about 0.3 and 0.4 mag respectively (Scoville et
al. 2000; see Fig. \ref{Genevamodels1}). For the extreme case A$_{V}$=30, the
corresponding shifts will be 0.6 and 0.8 mag, i.e. in region C1 the F160W-F222M color will move from the observed 2.67 to 2.07. Thus the colors modeled with this alternative dust geometry would imply that a hot dust contribution is not mandatory in the external less extincted regions, but it is still required to explain the color excess, in the highly extincted regions (A, B1 and C1).

The stellar mass of the young population has been measured using two different indicators, the H$\alpha$ luminosity and the M$_{160W}$ absolute magnitude, and comparing with the predicted values for the ages derived using the EW(H$\alpha$). In both cases the H$\alpha$/H$\beta$ ratio has been used for correct by extinction (for A and C$'$ the Pa$\alpha$/H$\alpha$ ratio was used too).

\begin{figure*}
\epsscale{0.5}
\plotone{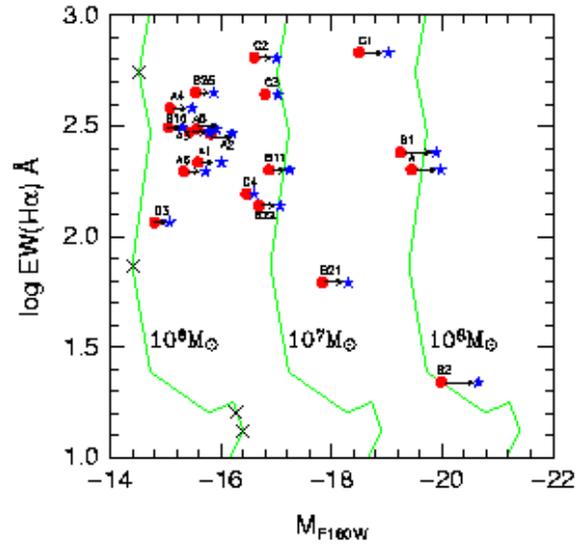}
\caption{Results of the stellar evolutionary
  models (SB99, Geneva High Isochrones, Salpeter IMF, using upper and lower mass boundaries of 0.1 and 100 M$_{\odot}$) compared with
   measured values. Long tracks are the output for
  instantaneous bursts as a function of age (from top to bottom, black crosses mark 4, 6, 8 and 10 Myr), and different cluster mass. The filled circles
  represents measured absolute F160W magnitude, and the stars represent the extinction corrected values from the H$\alpha$/H$\beta$ ratio. [\textit{See the electronic edition of the Journal for a color version of this figure.}]}\label{Genevamodels2}
\end{figure*}

The results obtained using M$_{160W}$ (see Fig. \ref{Genevamodels2} and Table
\ref{tablamodic}) show that the regions with more young stellar mass content
are nucleus A and regions B2, B21 and C1 (460, 300, 58 and 81$\times$10$^{6}$
M$_{\odot}$ respectively). The stellar mass content for the rest of the
regions, including the low-surface brightness knots, ranges between 3.1 and
14.0$\times$10$^{6}$M$_{\odot}$ (details on Table \ref{tablamodic}). Previous
derivations of the stellar mass of the clusters using absolute H-band
magnitudes vs H-K (not corrected for extinction) with a different stellar
populations synthesis model range, from about 1.5$\times$10$^{5}$M$_{\odot}$ to 5.0$\times$10$^{6}$M$_{\odot}$ (AAH00).

Using the H$\alpha$ luminosity, the young stellar mass content is again concentrated on the nucleus A, and regions B2, B21 and C1 (655, 679, 74 and 74$\times$10$^{6}$ M$_{\odot}$
respectively). For region C$'$, where there is not near-IR photometry, the mass derived using the H$\alpha$ luminosity is 115$\times$10$^{6}$M$_{\odot}$. The rest of the derived masses range from 5.7 to 57$\times$10$^{6}$ M$_{\odot}$. Comparing the mass in newly formed stars with the derivations of AAH00, there are no significant differences except for B2 where we derive a value two times higher. The masses of the young H$\alpha$-emitting regions identified in Arp~299 are within the range of masses founded in LIRGs (3$\times$10$^{6}$-10$^{8}$M$_{\odot}$, Scoville et al. 2000; Alonso-Herrero et al. 2002) and are on average about 100 times more massive than the galactic globular clusters (10$^{5}$M$_{\odot}$, van den Bergh 1995), probably due to the fact that the present regions have conglomerates of less massive clusters.

When comparing the stellar masses derived with the H$\alpha$ luminosity with the mass obtained from the absolute F160W magnitude, there is a good agreement in the nucleus A, but for the rest of the regions there are differences from a factor $\sim$1 to 5. One of
the main reasons for this difference is that in the majority of the cases (except A and B2) an aperture of 0\farcs5 has been used for the HST cluster
photometry, instead the 0\farcs9 or 1\farcs0 used for the INTEGRAL data
analysis (for the EW(H$\alpha$) and L(H$\alpha$) estimates); this difference can lead to offsets of a factor 3 in the mass estimates. Factoring in the aperture effects, the two methods give the same mean value within a factor 2.

\subsection{Contribution to the bolometric luminosity}
The derived ages of the young H$\alpha$-emitting star-forming regions cover a range of between 3$\times$10$^{6}$ years to at least 7-8$\times$10$^{6}$ years old. This is consistent with the methodology used here: since the H$\alpha$ traces preferentially the massive young ionizing stellar clusters, population older than 10 Myr is difficult to be detected as H$\alpha$ decreases. In any case there is further evidence that the star formation in Arp~299 has persisted for at least the last 10-15$\times$10$^{6}$ years (AAH00), with high activity on the star-forming regions. On the basis of this dispersion, we consider that the overall star formation of Arp~299 can be modeled by a continuous star formation. Under this premise the star formation rate (SFR) and the IR luminosity contribution can be derived from the extinction corrected H$\alpha$ luminosity, using Kennicutt (1998) expressions, valid for a Salpeter IMF with mass cutoffs of 0.1-100 M$_{\odot}$:

\begin{equation}
SFR(M_{\odot}~year^{-1})=7.9 \times 10^{-42} L(H\alpha) (erg~s^{-1})
\end{equation} 

\begin{equation}
L_{IR} (erg~s^{-1})=175.5 \times L(H\alpha) (erg~s^{-1})
\end{equation}

that strictly speaking are only valid for starbursts with ages less than 10$^{8}$ years.

Without taking into account the nucleus B1 (AGN candidate), the SFR of the
entire system is about 43 M$\odot$ year$^{-1}$, and the L$_{IR}$ is
9.5$\times$10$^{44}$ erg s$^{-1}$, whereas the luminosity derived from the
IRAS fluxes is 2.2$\times$10$^{45}$ erg s$^{-1}$, thus the contribution of the studied regions to the IR luminosity is about 45\%. There is a lack of about 55\% in luminosity, that can be accounted for by the AGN located in B1. Previous
estimates based on the mid and far-IR spectra of Arp~299 assign a
 20\% of the L$_{IR}$ coming from B1 and B2 (Charmandaris et
 al. 2002), while at 11.7 $\mu$m the percentage of flux coming from B1 and B2 is about 30\% (Keto et al. 1997). So even considering that the majority of this contribution comes from B1, there is  still a lack of about 25$-$35\%. This inconsistence is probably due to the fact that the present result is limited by the derived extinction and/or the models.

\section{2D Gas kinematics and dynamical mass derivations}\label{sec-kin}

Previous studies of Arp~299 (Gehrz et al. 1983; Telesco
et al. 1985) concluded that this system consists of two spiral galaxies disturbed by the merging process, which are rotating in opposite sense (Augarde \& Lequeux 1985). More recently Hibbard \& Yun (1999) have described Arp~299 as a structure formed after a prograde-retrograde encounter between IC~694 (Sab-Sb) and NGC~3690 (Sbc-Sc) that took place
some 750 Myr ago, and having a long tidal tail (180 kpc in length) detected in H\,{\sc{i}}. Although there are still two structures clearly distinguishable with the two nuclei separated by about 4.6 kpc, the warm gas component has lost its kinematic identity and
the velocity field presents a very irregular pattern (Rampazzo et al. 2005). 

\subsection{Gas kinematics}\label{seckin}
The kinematics of the ionized and the neutral gas is traced  by the H$\alpha$ emission line and the  Na\,{\sc i}$\lambda$5890 absorption line respectively (Sparks et al. 1997). 
The ionized gas mean velocities and velocity dispersions were obtained by
fitting the H$\alpha$+[N\,{\sc ii}]$\lambda\lambda$6548,6584 complex with
a single Gaussian velocity component per line; the same method was applied
to the He\,{\sc i}$\lambda$5876+Na\,{\sc i}$\lambda\lambda$5890, 5896 complex. 
The values derived for the individual regions are presented in Tables
\ref{kinearp299}, \ref{kineic694} and \ref{kinengc3690}. For the
velocity measurements, the errors have been calculated taking into
account both the systematic uncertainties of the wavelength
calibration (same for all lines) in the velocity measurement, and the
uncertainties in the measurement of the centroid of the line (that
depends on the S/N of the studied line). One of the most relevant
results obtained is that the two dimensional structure of the ionized
and neutral gas velocity field show similar features in both the
individual galaxies and the interface region, indicating that the
ionized and neutral gas kinematics are coupled.

 The structure of the H$\alpha$ velocity dispersion shows that there is a variation on the average velocity dispersion values: 
the interface region has the lowest (45 km s$^{-1}$), IC~694 the intermediate (65 km s$^{-1}$) and NGC~3690 (120 km s$^{-1}$) the maximum values of $\sigma$.

\subsubsection{Interface Region}\label{kinematinter}
\begin{figure*}
\epsscale{1.0}
\plotone{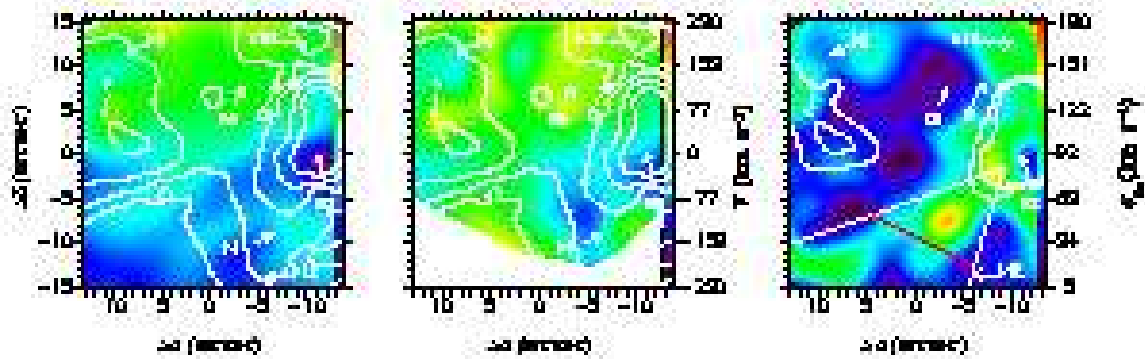}
\caption{Arp~299 interface region kinematics. Velocity field of the
  ionized gas as traced by the H$\alpha$ line (\textit{Left panel}), neutral
  gas as traced by the Na\, {\sc i} line (\textit{Center panel}) and velocity
  dispersion of the ionized gas (\textit{Right panel}). The velocities are
  referred to the nucleus A as observed with the higher spatial resolution SB2
  bundle, i.e. 3121$\pm$27 km s$^{-1}$ for the ionized and 3057$\pm$34 km
  s$^{-1}$ for the neutral gas. Lines depicted the ionization cone as in Fig \ref{Diaginterface}. [\textit{See the electronic edition of the Journal for a color version of this figure.}]}\label{kinematinterface}
\vspace{0.4cm}
\end{figure*}

The overall velocity field of the ionized gas as observed with the SB3
bundle has a complex and irregular
pattern, that does not indicate the presence of virialized simple (e.g.,
rotating) motions, on scales of 0.6 kpc or larger
(Fig. \ref{kinematinterface}, \textit{Left}). The structure of this
map agrees well with that obtained by Rampazzo et al. (2005) covering
a larger area (FoV of 5\arcmin.9) on the sky. The velocities in the northern
part are receding, and in the southern part are approaching to the systemic
value, defined as the velocity in nucleus A (the dust-enshrouded nucleus of IC~694)
 with the higher spatial resolution bundle SB2, V$_{A}$=3121$\pm$27 km s$^{-1}$ (see
section \S \ref{seckineic694}). The corresponding lower spatial
resolution SB3 velocity is V$_{A}$=3190$\pm$27 km s$^{-1}$. 

The velocity field of the neutral gas obtained from the NaD absorption lines, 
despite of being noisier (the intensity of the NaD can be typically 2\% of that 
of H$\alpha$), shares similar structural characteristics with the ionized gas velocity field (Figure \ref{kinematinterface} 
\textit{Center panel}), and presents a systemic velocity of 3057$\pm$34 km s$^{-1}$.

 The ionized gas velocity field presents a peak-to-peak velocity of about 230 km s$^{-1}$, while for the neutral gas the variation is smaller, about 170 km~s$^{-1}$. In both cases the area with the largest approaching velocities is
close to B2. 
The ionization cone does not show relevant features in the ionized gas velocity field, nor in the neutral gas velocity field.
With the exception of the nuclear regions, the ionized gas in the
galaxies  does not have a distinct kinematics with respect to the interface
region (i.e., the structure of the velocity field does not correlate with the
distribution of mass as traced by the continuum images). 

The velocity gradient of the ionized gas through the interface
is about 40 km s$^{-1}$ kpc$^{-1}$, compatible with the H$\alpha$
velocity gradients previously measured across ULIRGs observed with a 
slit (up to 47 km s$^{-1}$ kpc$^{-1}$ for galaxies with double nuclei, 
Martin 2006).

The velocity dispersion shows distinct features over the entire area; there
are large differences in $\sigma$, that allow us to discriminate spatially
between quiet and turbulent gas with macroscopic motions
(Fig. \ref{kinematinterface} \textit{Right panel}). The so-called
interface has an average $\sigma$ value of 45 km s$^{-1}$ that
traces no turbulent gas. The low velocity amplitude and velocity
dispersion of the interface indicates that the ionized gas could be in
a close to quiescent phase or slowly rotating.
 Larger $\sigma$ values are measured on the
previously defined ionization cone, with the maximum (160$\pm$21 km
s$^{-1}$) located on the peak of Seyfert-like ionization. This spatial coincidence 
could indicate that in that area the two contributions of the biconic surface of the cone are added, and thus the line measured is wider (Heckman et al. 1990).
On the other hand, the maximum velocity dispersion would imply that
the largest macroscopic motions, with approaching velocities, are
taking place in that highly ionized area. 

\subsubsection{IC 694}\label{seckineic694}
\begin{figure*}
\epsscale{1.0}
\plotone{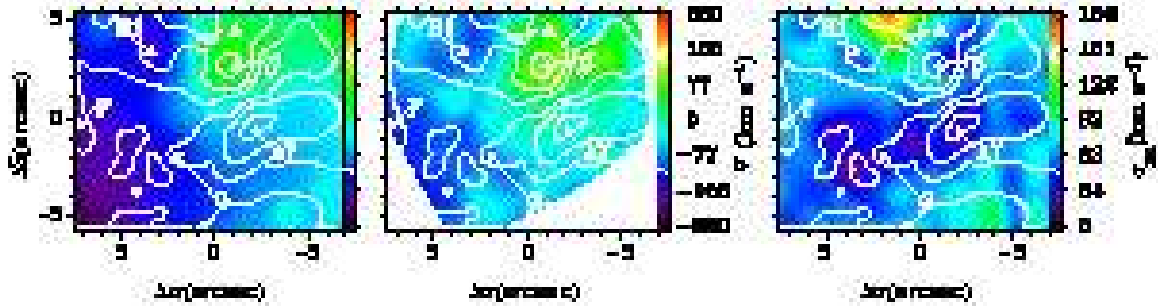}
\caption{IC~694 kinematics. Velocity field of the
  ionized gas as traced by the H$\alpha$ line (\textit{Left panel}), neutral
  gas as traced by the Na\, {\sc i} line (\textit{Center panel}) and velocity
  dispersion of the  ionized gas (\textit{Right panel}). The velocities are
  referred to the nucleus A,  for which the ionized gas velocity is 3121$\pm$27 km
  s$^{-1}$ and the  neutral gas velocity is 3057$\pm$34 km s$^{-1}$. [\textit{See the electronic edition of the Journal for a color version of this figure.}]}\label{kinematic694}
\vspace{0.4cm}
\end{figure*}

The velocity field of both the ionized and neutral gas presents a
similar structure (Fig. \ref{kinematic694} \textit{Left and Center panels}), when
observed with the SB2 bundle. It has an
irregular  structure, with the largest receding velocities associated with
region I (3181$\pm$27 km s$^{-1}$ for the ionized gas and 3155$\pm$34
km s$^{-1}$ for the neutral), located at 0.42 kpc SW from the nucleus (A). These maps show a general trend with velocities
changing from NW (receding) to SE (approaching), in agreement with the pattern
found with the SB3 data in this region.
 The systemic velocity for the ionized gas is defined as the velocity
 associated with the nucleus (region A), and it has a value of 3121$\pm$27
 km~s$^{-1}$, consistent with Zhao et al. (1997) estimate based on the
 H92$\alpha$ radio recombination line (3120$\pm$50~km~s$^{-1}$). The ionized
 gas velocity field has a peak-to-peak variation of 290 km~s$^{-1}$.  The neutral gas systemic velocity is 3057$\pm$34 km
s$^{-1}$, and the peak-to-peak velocity is about 216 km s$^{-1}$. Although
 evident structural similarities are found between the ionized and neutral gas (i.e. same kinematics), in general the velocity of the neutral gas presents a
 blueshift over the entire area with respect to the ionized gas (about
 -30$\pm$12 km s$^{-1}$ over the studied regions). This systematic blueshift
 appears to be real taking into account the uncertainties. In that case, this
 is a marginal effect,  with a lower value than the ones measured in ULIRGs using a slit (Martin 2006; from about -60 km s$^{-1}$ to -300 km s$^{-1}$). In the present analysis the value obtained is derived from IFS data, i.e. has the advantage of the two dimensional distribution versus the use of the slit, where the measurements are limited to a restricted area, probably centered on the nuclear regions.

Across the galaxy there are gradients of about 110 km s$^{-1}$
kpc$^{-1}$ in the ionized gas velocity (between regions located $\sim$2.3 kpc apart i.e. $\sim$11\arcsec), higher than the typical values founded in ULIRGs (up to 60 km s$^{-1}$ kpc$^{-1}$ for an individual galaxy, Martin 2006) when using a slit of 20\arcsec\, long. In closer regions with large velocity variations the gradients can be 150 km s$^{-1}$ kpc$^{-1}$ (between regions I and northern of G, located about 1.4 kpc apart). In spite of these gradients, and as discussed below, there are no clear evidence for virialized simple rotating motions.

The two-dimensional velocity dispersion map shows a different structure than the velocity field (Fig. \ref{kinematic694},
\textit{Right panel}). For nucleus A, the $\sigma$ value is 108$\pm$11 km s$^{-1}$,
in well agreement with the derivation of Shier et al. (1996) based on the
CO(2,0) near-infrared band at 2.3 $\mu$m (135$\pm$21 km s$^{-1}$). The maximum value for $\sigma$ (170$\pm$20 km s$^{-1}$) is located in a region 0.28 kpc towards the north-east of A, where the [O\, {\sc{i}}]$\lambda$6300 emission peak is also located. Since [O\, {\sc{i}}]$\lambda$6300 trace shocks, this high $\sigma$ value is a hint of turbulence and shocks. The lowest velocity
dispersion (lower limit 25$\pm$3 km s$^{-1}$) corresponds to the area dominated
by H\,{\sc ii}-like ionization, towards the east of the spiral arm (e.g. A7,
see \S \ref{ionizic}). This indicates that the velocity dispersion is
likely dominated by the internal kinematics within the stellar clusters, rather than by large regions projected along the line of sight.

In this case it is not possible to state that there are clear indication of pure rotation, i.e. the photometric peak does not coincide with the velocity dispersion maximum, and neither of them coincide with the possible kinematic center. The velocity maps show that even if pure rotation does not dominate the galaxy kinematics,  rotational movements can be present. 

\subsubsection{NGC~3690}\label{kinngc}

\begin{figure*}
\epsscale{1.0}
\plotone{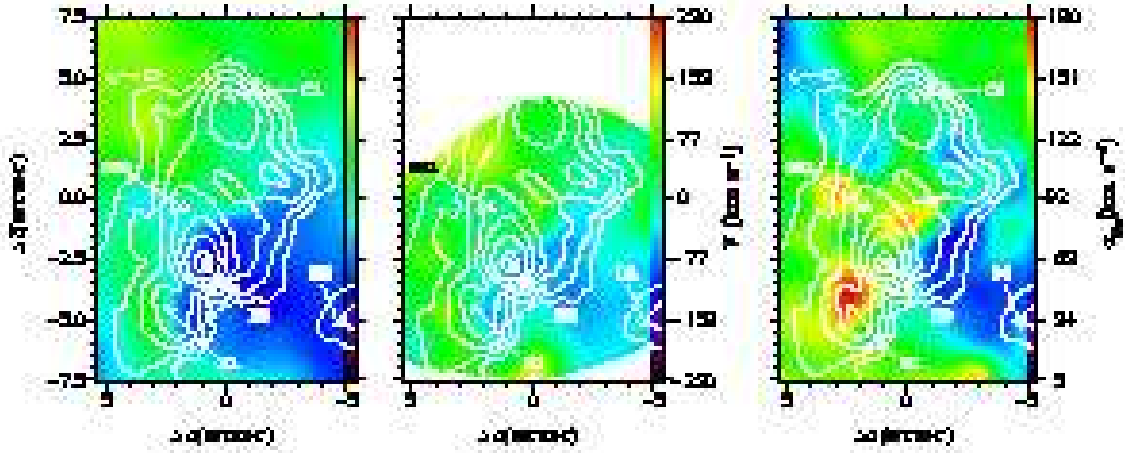}
\caption{NGC~3690 kinematics. Velocity field of the
  ionized gas as traced by the H$\alpha$ line (\textit{Left panel}), neutral gas as traced
  by the Na\, {\sc i} line (\textit{Center panel}) and velocity dispersion of the
  ionized gas (\textit{Right panel}). The velocities are
  referred to the nucleus B1,  for which the ionized gas velocity is 3040$\pm$27 km
  s$^{-1}$ and the  neutral gas velocity is 2976$\pm$32 km s$^{-1}$. Note that the nucleus B1 presents the the highest $\sigma$ value. [\textit{See the electronic edition of the Journal for a color version of this figure.}]}\label{kinematngc3690} 
\vspace{0.4cm}
\end{figure*}

Like in the previous case, at a resolution of 0.18 kpc the
structure of the ionized and neutral gas velocity fields obtained with the SB2
bundle are very similar (Fig. \ref{kinematngc3690} \textit{Left and Center panels}).
The largest receding velocities are detected towards the north-east of the
field (C$'$, C6), while the largest approaching velocities are located on B2 and
towards the south-west. The ionized gas systemic velocity is defined as the one
of the nucleus B1, with a value of 3040$\pm$27 km s$^{-1}$. This value coincides with Zhao et al. (1997) measurements using the
H92$\alpha$ radio recombination line (3080$\pm$40 km s$^{-1}$). For the
ionized gas, the peak-to-peak velocity  is 257 km s$^{-1}$.
The systemic velocity measured with the neutral gas is 2976$\pm$32km s$^{-1}$, with a peak-to-peak variation of 217 km s$^{-1}$. In both ionized and neutral gas,
the velocity variations are similar to the ones measured in IC~694. As founded in ULIRGs (Martin 2006), there is a systemic blueshift of the neutral gas velocity with respect to the ionized gas (about -24$\pm$13km s$^{-1}$ in the studied regions). Due to its value, lower than the results founded for ULIRGs (from about -60 km s$^{-1}$ to -300 km s$^{-1}$, Martin 2006), this blueshift is consider as a marginal effect.

Across the galaxy there are gradients of about 70 km s$^{-1}$
kpc$^{-1}$ in the ionized gas (between C$'$ and D3). This result is close to the gradients derived for ULIRGs using a slit (up to 60 km s$^{-1}$ kpc$^{-1}$ for an individual galaxy, Martin 2006).

The two-dimensional structure of the velocity dispersion ranges between
50$\pm$5 and 172$\pm$20 km s$^{-1}$ (Fig. \ref{kinematngc3690}
\textit{Right panel}), and present different characteristics than the
two dimensional velocity fields. The nucleus B1 has the maximum
velocity dispersion value of the entire system (172$\pm$20 km
s$^{-1}$), and regions with high velocity dispersion ($\sim$160 km
s$^{-1}$) are also detected about 1 kpc towards the north and
north-west of B1. Looking at the ionized gas velocity field, this
secondary peaks of the velocity dispersion are located in regions
where there are high velocity variations (from approaching to
receding), and thus this $\sigma$ would represent a convolution of a
large velocity gradient. Although the $\sigma$-peak coincides
spatially with the nucleus, and then it would be tracing the mass
(large $\sigma$ implies that this is the most massive galaxy of the system), the presence of an AGN in B1 could produce a broadening of the line width.

Like in the case of IC~694 there are no clear evidences for ordered rotating motions; the photometric and velocity dispersion peaks do not coincide with the possible kinematic center. In conclusion, although the pure rotation does not dominate the kinematics of the galaxy, rotational process would be present.

\subsection{Where is the true nucleus of NGC~3690?}\label{truenucleus}
The galaxy NGC~3690, shows variations on the surface brightness importance of the regions B1 and B2 with wavelength. While in the ultraviolet (Windhorst et al. 2002) and in the optical (this work) B2 appears brightest than B1, and therefore has been identified as the nucleus, the reverse is true in the near-IR (2.2 $\mu$m AAH00) and mid-IR (Gallais et al. 2004). Besides, the optical kinematical analysis shows that the $\sigma$
 peak is located in B1 (see \S\ref{kinngc}). This fact along with the high extinction (A$_{V}$=3.7), the hot dust component needed to explain the color excess, and the presence of an AGN, lead us to the conclusion that B1 is the true nucleus of the individual galaxy NGC~3690, as previously discussed by Gallais et al. (2004) with mid-IR spectro-imaging, and concluded by Zezas et al. (2003) using Chandra X-ray observations.

\subsection{Dynamical mass derivation of the nuclei and brightest regions}\label{dynmass}
Until now the only mass derivation in this work has been done for the H$\alpha$-emitting young star forming regions. The central ionized gas velocity dispersion is a good tracer of the dynamical mass, thus it can be used as an alternative method for mass derivation. Assuming virialization, the dynamical mass of the nuclei and brightest regions can be derived as:

\begin{equation}
M_{dyn}=1.75\times10^{3} R_{hm} \sigma^{2} M_{\odot}
\end{equation}
where $R_{hm}$ is the half mass radius in pc and $\sigma$ is the velocity
dispersion in km s$^{-1}$. Assuming a constant value of 1.75 (see Colina et
al. 2005 for a detailed discussion), and that the effective radius (R$_{eff}$) of the light distribution traces the half mass radius in each region, it is possible to infer the total dynamical mass. 

In the present treatment we assume that H$\alpha$ is traced by Pa$\alpha$, that gives the real size of the emitting regions. In consequence the effective radius of the most important emitting sources is calculated over the HST Pa$\alpha$ image; additional estimates using the F110M image are done for comparison.
There are some caveats related to the spatial resolution of the different set of data worth mentioning. For this
mass estimate we are dealing with information at different spatial
resolutions: on the one hand the velocity dispersion from the
INTEGRAL SB2 bundle data was measured at a 0\farcs9 ($\sim$180 pc) resolution. On the
other hand the R$_{eff}$ over the most relevant Arp~299 regions (i.e. the ones
resolved) was measured 
using the NICMOS1/F110M image (0\farcs043/pix, i.e. $\sim$9 pc/pix) and
NICMOS2/Pa$\alpha$ image (0\farcs076/pix, i.e. $\sim$16 pc/pix). The typical size of the H\,{\sc{ii}} regions in LIRGs is 200-300 pc (Alonso-Herrero et al. 2002), i.e. several times larger than the effective radius of the stellar clusters and about the size of a SB2 fiber. Taking this into account, the clusters will be point sources for INTEGRAL, and their $\sigma$ will be integrated on the fiber.

For nucleus A we derive a dynamical mass of about 1.2$\times$10$^9$M$_{\odot}$
(Table \ref{tablamasic}) which is about 5 times smaller than the derivation
done by Shier et al. (1996) using the velocity dispersion measured from the
2.3$\mu$m CO band; this discrepancy is due to a different estimate of the
radius (r$_{CO}$/r$_{Pa\alpha}$$\simeq$4). The calculated M$_{dyn}$ implies
that the young stellar mass content derived from L(H$\alpha$) in A is
about 40\%. The mean value for the dynamical mass of nucleus B1 is
1.5$\times$10$^9$M$_{\odot}$ (in good agreement with Shier et al. 1996),
consequently the upper limit for the young stellar mass content is about 15\%. For B2 the mean
value  of the total mass is 0.6$\times$10$^9$M$_{\odot}$ (the same as
Shier et al. 1996). This would imply that about 100\% of the mass in B2 is from young stellar population, indicating that the mass derivation done using L(H$\alpha$) is probably overestimated. For the last two regions (C1 and C$'$), the R$_{eff}$ gives an upper limit for
the dynamical mass (0.6 and 0.3$\times$10$^9$M$_{\odot}$ respectively, the
latter according to the $>$0.1$\times$10$^9$M$_{\odot}$ from AAH00), that
implies young stellar mass contents of about 13\% and 40\% respectively.

\section{Summary}\label{sec11}
A detailed IFS study of the LIRG Arp~299 (IC~694+NGC~3690) system covering a region of
7.0$\times$6.2 kpc has been presented. This work illustrates the
flexibility and power of the IFS in studying  in two dimensions and
simultaneously the different physical properties, such as stellar populations,
internal dust extinction, ionization structure and gas kinematics, of complex
systems like Arp~299. In what follows only the main results of this study are summarized:

\begin{enumerate} 
\item The observed overall two dimensional stellar and ionized gas structure
differs due to internal extinction effects caused by the dust, and to the
different spatial distribution of the diverse (AGN, stars, shocks) ionizing
sources. The stellar continuum and emission line distribution peaks are not
spatially coincident but separated up to 1.4 kpc. The peaks of the ionized gas
traced by different emission lines (e.g. hydrogen recombination, high
excitation [OIII], and shock-tracer [OI]) are also not coincident indicating
that the ionizing mechanisms change on scales of about 200 pc or less.

\item The region B1 is identified as the true nucleus of NGC~3690. The presence of an AGN in B1 is inferred from the detection of a high-excitation Seyfert-like conical structure of about 1.5 kpc in size, an opening angle  of about 54\,$^{\circ}$, and its apex
being located in B1. This region also presents the additional characteristics
of a dust -enshrouded nucleus with high extinction (A$_{V}$=3.7$\pm$0.6), red colors (F160W-F222M=2.06$\pm$0.1) indicating hot dust emission, and high velocity dispersion ($\sigma$=172$\pm$20 km s$^{-1}$). 

\item The observed near-IR colors reveal the presence of hot dust over the entire structure. According to he F160W-F222M color, the upper limit for the mean hot dust contribution is about 30$\%$ at 600 K for the majority of the star-forming regions. The nuclei A and B1, and region C1 are the most affected by the hot dust component, with an upper limit of about 70$\%$, 80$\%$ and 90$\%$ at 600 K respectively.

\item The high surface brightness regions are mainly dominated by star
  formation, while the low surface brightness regions are dominated by
  LINER-like ionization process. The H$\alpha$-emitting star-forming regions
  present an extinction of between A$_{V}$=1$-$6 mag, with ages between 3.3
  and 7.2$\times$10$^{6}$ years, and masses that range between about 6 and 680$\times$10$^{6}$ M$_{\odot}$. The total contribution (extinction corrected) of these sources to the bolometric luminosity is about 45\%.

\item The structure of the ionized (H$\alpha$) and neutral (NaD) gas two dimensional velocity fields show similar features, indicating that both components are kinematically coupled on scales of hundreds of pc and larger. The velocity fields are very complex, and in general the galaxy does not appear to be dominated by ordered virialized simple (e.g. rotation) motions.  The morphology of
  the velocity field does not correlate with the distribution of mass as
  traced by the continuum images, that is, no spatial coincidence
  between the kinematic and the photometric centers exists. 

On average, the interface region presents the lowest value of velocity dispersion ($\sigma$$\sim$45 km s$^{-1}$), IC~694 presents the intermediate values ($\sigma$$\sim$65 km s$^{-1}$), and NGC~3690 the highest ($\sigma$$\sim$120 km s$^{-1}$). This indicate that NGC~3690 is the warmest body of the entire system, and probably the most massive. On the contrary, the low velocity amplitude and velocity dispersion of the interface suggests that the ionized gas is slowly rotating or in a close quiescent phase.

The velocity dispersion measurements over the nuclei and the brightest emitting regions,  imply a mean value for the dynamical mass of about 1.2$\times$10$^{9}$ M$_{\odot}$ and 1.5 $\times$10$^{9}$ for nuclei A and B1 respectively, within a mean radius of 58 pc and 28 pc. For regions B2, C1 and C$'$ the mass ranges between $\sim$0.2$\times$10$^{9}$ M$_{\odot}$ and $\sim$0.6$\times$10$^{9}$ M$_{\odot}$, within a radius of about 30 pc.

\end{enumerate}

\smallskip
\textbf{Acknowledgments}\\
The authors wish to thank Dr. Hajime Sugai, the referee, for his detailed and
constructive report. We thank Jes\'us Ma\'{\i}z-Apell\'aniz for his help using
\textit{jmaplot}. This work has been supported by the Spanish Ministerio de
Educaci\'on y Ciencia, under grant BES-2003-0852, projects AYA2002-01055 \& ESP2005-01480.

\clearpage

\begin{deluxetable}{lccccccccc}
\tabletypesize{\scriptsize}
\tablecolumns{10}
\tablecaption{Properties of the regions of the IC~694+NGC~3690 interface}
\tablehead{\colhead{Reg} & \colhead{Dist\tablenotemark{a}}&
    \colhead{$\Delta\alpha$\tablenotemark{a}} & \colhead{$\Delta\delta$\tablenotemark{a}} 
& \colhead{F$_{obs}$(H$\alpha$)\tablenotemark{b}} &
\colhead{A$_{V}$\tablenotemark{c}} 
& \colhead{log([OIII]/H$\beta$)\tablenotemark{d}} &
\colhead{log([OI]/H$\alpha$)\tablenotemark{e}} &
\colhead{log([NII]/H$\alpha$)\tablenotemark{f}} &
\colhead{log([SII]/H$\alpha$)\tablenotemark{g}}\\
\colhead{} & \colhead{(kpc)} & \colhead{(arcsec)} &\colhead{(arcsec)} &
\colhead{} &
\colhead{(mag)} & \colhead{} &
\colhead{} & \colhead{} & \colhead{}}
\startdata
K1  & 2.97 & -17.6 & -12.7 & 1.7 & 2.2 & 0.27  &-1.07 &-0.34 &-0.37\\
K2  & 3.81 & -21.6 & -13.7 & 1.1 & 0.9 & 0.07  &-1.04 &-0.36 &-0.38\\
K3  & 1.28 & -9.6  & -10.7 & 0.6 & 2.0 & 0.22  &-0.59 &-0.21 &-0.25\\
K4  & 2.49 & -13.0 & -1.8  & 2.7 & 1.6 & 0.04  &-0.83 &-0.41 &-0.31\\
K5  & 3.53 & -14.5 &  3.9  & 10.7& 2.9 & 0.14  &-1.44 &-0.54 &-0.64\\
K6  & 3.71 & -13.1 &  6.0  & 5.7 & 2.3 & -0.05 &-1.27 &-0.48 &-0.54\\
K7  & 2.72 & -10.6 &  1.9  & 4.0 & 2.0 & 0.01  &-1.10 &-0.47 &-0.45\\
K8  & 4.07 & -9.0  &  9.8  & 1.4 & 2.7 & 0.14  &-0.71 &-0.32 &-0.23\\
K9  & 3.52 & -2.6  &  7.8  & 3.3 & 1.8 & -0.01 &-0.92 &-0.37 &-0.36\\
K10 & 5.26 & -21.5 & -9.0  & 4.2 & 1.0 & 0.17  &-1.36 &-0.53 &-0.56\\
K11 & 1.44 & -9.1  & -5.1  & 1.6 & 1.5 & 0.02  &-0.82 &-0.39 &-0.31\\
K12 & 3.18 & -16.8 & -17.4 & 7.6 & 2.0 & 0.13  &-0.98 &-0.35 &-0.37
\enddata
\tablecomments{\footnotesize (a): Distance relative 
to the nucleus A. (b): Flux measured ($\times$10$^{-14}$erg s$^{-1}$cm$^{-2}$) using a circular aperture of 3\farcs\, Uncertainties 10\%-15\%. (c): Internal extinction derived from the
  H$\alpha$/H$\beta$ ratio. Average error in A$_{V}$ is 0.7
  magnitudes. (d): Average error $\pm$0.11 (e): Average error
  $\pm$0.06 (f): Average error $\pm$0.05 (g): Average error $\pm$0.06. In all cases, emission
  line ratios are derived using extinction corrected flux values.}\label{infoarp}
                               
\end{deluxetable}

\clearpage

\begin{deluxetable}{lccccccccc}
\tabletypesize{\scriptsize}
\tablecolumns{10}
\tablecaption{Properties of the regions of IC~694}
\tablehead{\colhead{Reg} & \colhead{Dist\tablenotemark{a}}&
    \colhead{$\Delta\alpha$\tablenotemark{a}} & \colhead{$\Delta\delta$\tablenotemark{a}} 
& \colhead{F$_{obs}$(H$\alpha$)\tablenotemark{b}} &
\colhead{A$_{V}$\tablenotemark{c}} 
& \colhead{log([OIII]/H$\beta$)\tablenotemark{d}} &
\colhead{log([OI]/H$\alpha$)\tablenotemark{e}} &
\colhead{log([NII]/H$\alpha$)\tablenotemark{f}} &
\colhead{log([SII]/H$\alpha$)\tablenotemark{g}}\\
\colhead{} & \colhead{(kpc)} & \colhead{(arcsec)} &\colhead{(arcsec)} &
\colhead{} &
\colhead{(mag)} & \colhead{} &
\colhead{} & \colhead{} & \colhead{}}
\startdata
A  & 0.00 &  0.0  & 0.0 & 2.9 & 2.9 & -0.14 & -1.03 & -0.31 & -0.41\\
A1 & 0.46 &  1.5  &-1.6 & 2.6 & 2.3 & -0.22 & -1.29 & -0.40 & -0.50\\
A2 & 0.54 & -1.0  &-2.4 & 3.1 & 2.0 & -0.14 & -1.51 & -0.52 & -0.60\\
A3 & 1.37 &  3.3  &-5.7 & 2.3 & 1.9 & -0.03 & -1.61 & -0.53 & -0.59\\
A4 & 1.51 &  3.8  &-6.3 & 3.0 & 2.1 & -0.02 & -1.73 & -0.55 & -0.66\\
A5 & 1.06 & -2.1  &-4.7 & 6.0 & 1.9 & 0.10  & -1.56 & -0.54 & -0.62\\
A6 & 1.00 & -3.6  &-3.9 & 4.0 & 2.1 & 0.04  & -1.46 & -0.52 & -0.56\\
A7 & 0.99 & -2.8  &-3.9 & 6.2 & 2.3 & 0.09  & -1.61 & -0.56 & -0.66\\
F  & 1.58 &  6.5  &-4.1 & 6.7 & 3.2 & 0.10  & -2.06 & -0.52 & -0.92\\
G  & 1.10 &  1.1  &-5.2 & 6.2 & 1.9 & 0.05  & -1.84 & -0.54 & -0.74\\
H1 & 0.48 &  2.3  &-0.3 & 6.4 & 2.8 & -0.15 & -1.41 & -0.47 & -0.63\\
H2 & 0.34 &  1.6  & 0.5 & 4.7 & 3.4 & -0.06 & -0.99 & -0.32 & -0.46\\
I  & 0.42 & -1.8  &-0.9 & 6.0 & 1.6 & -0.09 & -1.60 & -0.52 & -0.65\\
I1 & 1.45 & -7.0  & 0.8 & 0.8 & 2.4 & -0.10 & -1.28 & -0.43 & -0.42\\
\enddata
\vspace{-0.4cm}
\tablecomments{\footnotesize (a): Distance relative 
to the nucleus A. (b): Flux measured ($\times$10$^{-14}$erg s$^{-1}$cm$^{-2}$)
using a circular aperture of 1\farcs\, Uncertainties  10\% - 15\%. (c): Internal extinction derived from the
  H$\alpha$/H$\beta$ ratio. Error in A$_{V}$ is 0.5 magnitudes. (d):
  Average error $\pm$0.08 (e): Average error $\pm$0.04 (f): Average
  error $\pm$0.05 (g): Average error $\pm$0.04. In all cases, emission
  line ratios are derived using extinction corrected flux values.}\label{tabic} 
                               
\end{deluxetable}                                                           

\clearpage

\begin{deluxetable}{lccccccccc}
\tabletypesize{\scriptsize}
\tablecolumns{8}
\tablecaption{Properties of the regions of NGC~3690}
\tablehead{\colhead{Reg} & \colhead{Dist\tablenotemark{a}}&
    \colhead{$\Delta\alpha$\tablenotemark{a}} & \colhead{$\Delta\delta$\tablenotemark{a}} 
& \colhead{F$_{obs}$(H$\alpha$)\tablenotemark{b}} &
\colhead{A$_{V}$\tablenotemark{c}} 
& \colhead{log([OIII]/H$\beta$)\tablenotemark{d}} &
\colhead{log([OI]/H$\alpha$)\tablenotemark{e}} &
\colhead{log([NII]/H$\alpha$)\tablenotemark{f}} &
\colhead{log([SII]/H$\alpha$)\tablenotemark{g}}\\
\colhead{} & \colhead{(kpc)} & \colhead{(arcsec)} &\colhead{(arcsec)} &
\colhead{} &
\colhead{(mag)} & \colhead{} &
\colhead{} & \colhead{} & \colhead{}}
\startdata
B1      & 0.00 &  0.0  &  0.0 &9.7  &3.7 &0.55 &-1.07&-0.29&-0.54\\
B2      & 0.49 & -1.5  &  1.9 &4.0  &3.9 &0.43 &-1.37&-0.35&-0.74\\
B11     & 0.22 & 0.27  & -1.0 &5.7  &2.0 &0.10 &-1.37&-0.41&-0.52\\
B16     & 0.26 & -0.5  & -1.1 &10.6 &1.1 &-0.02 &-1.68&-0.47&-0.71\\
B21     & 0.82 & -1.0  &  3.9 &3.1  &2.6 &0.02 &-1.54&-0.48&-0.77\\
B22     & 0.63 & -2.9  &  1.0 &3.2  &2.1 &0.02 &-1.51&-0.41&-0.68\\
B25     & 0.85 & -3.9  &  1.3 &7.2  &1.7 &-0.02&-1.88&-0.47&-0.80\\
C1      & 1.83 & -2.4  &  8.6 &28.1 &3.0 &0.13 &-1.87&-0.49&-0.96\\
C2      & 1.80 & -3.0  &  8.2 &33.4 &2.1 &0.11 &-1.92&-0.46&-0.94\\
C3      & 1.75 & -3.9  &  7.6 &16.7 &1.0 &0.01 &-1.80&-0.44&-0.78\\
C4      & 1.39 & -4.2  &  5.3 &5.7  &0.5 &-0.03&-1.80&-0.47&-0.67\\
C6      & 1.55 &  1.0  &  7.5 &5.2  &1.9 &0.02 &-1.74&-0.53&-0.70\\
C$^{'}$ & 1.96 &  2.7  &  9.5 &0.9  &3.8 &0.06 &-1.40&-0.45&-0.57\\
D3      & 1.49 & -7.2  & -0.9 &1.3  &1.3 &-0.03&-1.60&-0.44&-0.55\\
D4      & 1.13 & -5.5  &  0.1 &2.6  &2.0 &-0.05&-1.44&-0.41&-0.57
\enddata
\tablecomments{\footnotesize (a): Distance relative 
to the nucleus B1. (b): Flux measured ($\times$10$^{-14}$erg
s$^{-1}$cm$^{-2}$) using a circular aperture of 0\farcs9. Uncertainties 10\% - 15\% ($\sim$25\% for B2).(c): Internal
extinction derived from the H$\alpha$/H$\beta$ ratio. Error in A$_{V}$ is 0.5
magnitudes (0.6 for B1 and C$'$, and $\sim$ 0.7 for B2). (d): Average error 
$\pm$0.13 (e): Average error $\pm$0.13 (f): Average error $\pm$0.05
(g): Average error $\pm$0.05. In all cases, emission line ratios
 are derived using  extinction corrected flux values.}\label{tabngc}
                                
\end{deluxetable}

\clearpage 

\begin{deluxetable}{lccc}
\tabletypesize{\footnotesize}
\tablewidth{0pc}
\tablecolumns{5}
\tablecaption{Pa$\alpha$/H$\alpha$-based internal extinction estimates}
\tablehead{\colhead{Region} &
\colhead{f(Pa$\alpha$)}& \colhead{f(Pa$\alpha$)/f(H$\alpha$)}& \colhead{A$_{V}$}\\
\colhead{} & \colhead{($\times$10$^{-15}$erg cm$^{-2}$s$^{-1}$)}& \colhead{}& \colhead{(mag)}}
\startdata
A   & 167  & 5.8 & 6.1\\
A1  & 2.74 & 0.1 & $-$\\
A2  & 5.68 & 0.2 & 0.9\\
A3  & 4.37 & 0.2 & 0.9\\
A4  & 9.62 & 0.3 & 1.5\\
A5  & 8.41 & 0.1 & $-$\\
A6  & 2.73 & 0.1 & $-$\\
A7  & 10.5 & 0.2 & 0.9\\
H1  & 12.6 & 0.2 & 0.9\\
H2  & 17.3 & 0.4 & 2.0\\
I   & 5.11 & 0.1 & $-$\\
I1  & $-$  & $-$ & $-$\\
B1  & 66.7 & 0.7 & 2.8\\
B2  & 6.10 & 0.2 & 0.9\\
B11 & 9.77 & 0.2 & 0.9\\
B16 & 18.5 & 0.2 & 0.9\\
B21 & 4.14 & 0.1 & $-$\\
B22 & 5.15 & 0.2 & 0.9\\
B25 & 11.0 & 0.2 & 0.9\\
C1  & 85.3 & 0.3 & 1.5\\
C2  & 76.3 & 0.2 & 0.9\\
C3  & 10.0 & 0.1 & $-$\\
C4  & 4.70 & 0.1 & $-$\\
C6  & 9.16 & 0.2 & 0.9\\
C$'$  & 38.1 & 4.1 & 5.6\\
D3  & 0.15 & 0.01 &$-$\\
D4  & 4.81 & 0.2 & 0.9
\enddata
\tablecomments{Fluxes of Pa$\alpha$ measured over the selected regions using
  the HST (F190N-F187N) image. Average uncertainty in A$_{V}$ is 0.7 magnitudes. Some values are discarded due to the noise affecting Pa$\alpha$
and H$\alpha$ low-surface brightness.}\label{Palfatable}
\end{deluxetable}   

\clearpage

\clearpage

\begin{deluxetable}{lccccc}
\tabletypesize{\footnotesize}
\tablewidth{0pc}
\tablecolumns{6}
\tablecaption{Observed magnitudes and colors}
\tablehead{\colhead{Region} &\colhead{m$_{F814W}$}&\colhead{m$_{F160W}$} &
  \colhead{F110M-F160W} & \colhead{F160W-F222M}& \colhead{F814W-F110M}}
\startdata
A   &16.85 &13.72  &1.33 &1.36   &1.80  \\
A1  &18.53 &17.59  &1.07 &1.04   &-0.13 \\
A2  &18.57 &17.36  &0.91 &0.57   &0.30  \\
A3  &18.40 &17.75  &0.81 &0.65   &-0.15 \\
A4  &18.84 &18.09  &1.05 &0.69   &-0.31 \\
A5  &18.19 &17.61  &0.72 &0.63   &-0.14 \\
A6  &17.87 &17.85  &0.70 &0.54   &-0.68 \\
B1  &15.67 &13.93  &1.23 &2.06   &0.51  \\
B11 &17.30 &16.31  &0.88 &0.64   &0.11  \\
B16 &16.90 &18.12  &0.19 &$-$     &-1.42\\
B2  &14.09 &13.19  &0.70 &0.69   &0.20  \\
B21 &16.16 &15.35  &0.79 &0.55   &0.03  \\
B22 &17.78 &16.49  &0.93 &0.75   &0.36  \\
B25 &18.01 &17.67  &0.94 &0.62   &-0.56 \\
C1  &17.04 &14.68  &1.71 &2.67   &0.66  \\
C2  &16.63 &16.57  &0.67 &$-$     &-0.61\\
C3  &17.04 &16.37  &0.70 &0.52   &-0.02 \\
C4  &17.24 &16.71  &0.82 &0.44   &-0.29 \\
D3  &18.74 &18.37  &0.70 &0.46   &-0.33 
\enddata
\tablecomments{Errors in the colors are $\pm$0.10 and $\pm$0.14 for the bright
  and faint sources respectively.}\label{Magtable} 
                               
\end{deluxetable}

\clearpage 

\begin{deluxetable}{lccccc}
\tabletypesize{\footnotesize}
\tablewidth{0pc}
\tablecolumns{5}
\tablecaption{Properties of the stellar populations in regions of
  IC~694 and NGC~3690}
\tablehead{\colhead{Region} &\colhead{L(H$\alpha$)\tablenotemark{a}}&\colhead{EW(H$\alpha$)} & \colhead{Age} &
  \colhead{Mass(L$_{H\alpha}$)\tablenotemark{b}} & \colhead{Mass(M$_{F160W}$)\tablenotemark{c}}\\
\colhead{} & \colhead{($\times$10$^{40}$erg s$^{-1}$)}& \colhead{(\AA)} 
& \colhead{($\times$10$^6$ years)} &
\colhead{($\times$10$^6$M$_{\odot}$)}& \colhead{($\times$10$^6$M$_{\odot}$)}}
\startdata
A        &9.2 (150)  &201  &5.3  &40.0 (655.0) &200.0 (460.0) \\
A1  &4.8        &217  &5.2  &17.6     &6.0       \\
A2  &4.6        &292  &5.0  &13.0       &6.4       \\
A3  &3.2        &297  &5.0  &9.0        &4.8       \\
A4  &4.7        &382  &4.8  &10.0       &4.0       \\
A5  &7.3        &307  &5.0  &21.0       &5.3       \\
A6  &6.1        &197  &5.3  &26.8     &5.0       \\
A7  &11.3       &277  &5.1  &36.5     &$-$       \\
F   &28.6       &760  &3.3  &21.8     &$-$       \\
G   &7.9        &654  &3.4  &6.2      &$-$       \\
H1  &18.6       &416  &4.7  &34.0       &$-$       \\
H2  &22.5       &314  &4.9  &57.0       &$-$       \\
I   &5.8        &220  &5.2  &21.5     &$-$       \\
I1      &1.6        &86      &5.9        &12.9        &$-$       \\
B1      &62.3$^*$   &240$^*$ & 5.2$^*$  &231.0$^*$    &200.0$^*$  \\
B2      &29.8       &22      & 7.2      &678.0        &300.0  \\
B11     &7.9        &200     & 5.3      &34.6         &14.0 \\
B16     &7.1        &311     & 5.0      &20.0         &3.1  \\
B21     &7.6        &62      & 6.1      &74.0         &58.0 \\
B22     &4.8        &138     & 5.6      &27.3         &10.0 \\
B25     &7.7        &449     & 4.6      &13.7         &5.6  \\
C1      &93.8       &680     & 3.4      &74.0         &81.0 \\
C2      &53.4       &645     & 3.5      &43.8         &9.8  \\
C3      &9.7        &440     & 4.7      &17.7         &9.8  \\
C4      &2.1        &156     & 5.4      &10.3         &8.4  \\
C6      &6.8        &609     & 3.6      &5.7          &$-$  \\
C$^{'}$ &5.2 (31.0) &245     & 5.2      &19.0 (115.0) &$-$  \\
D3      &1.0        &116     & 5.7  &6.4      &3.2  \\
D4      &3.6        &360     & 4.8  &7.8      &$-$  
\enddata
\tablecomments{(a) Luminosity corrected for extinction using
  H$\alpha$/H$\beta$ ratio; the extinction using Pa$\alpha$/H$\alpha$
  ratio is given for regions A and C$'$ in brackets. (b) Mass derived from the  H$\alpha$ luminosity corrected by extinction using H$\alpha$/H$\beta$ ratio, and assuming the age listed in column 4. In brackets estimate using the Pa$\alpha$/H$\alpha$ extinction-corrected data. (c) Mass derived from the absolute magnitude M$_{F160W}$ corrected by extinction using H$\alpha$/H$\beta$ ratio, and assuming age listed in column 4. In brackets estimate using the Pa$\alpha$/H$\alpha$ extinction-corrected data. $^*$ means AGN contribution.}\label{tablamodic} 
                               
\end{deluxetable}

\clearpage
\begin{deluxetable}{lccc}
\tabletypesize{\footnotesize}
\tablewidth{0pc}
\tablecolumns{4}
\tablecaption{Kinematic of the interface region}
\tablehead{\colhead{Region} & \colhead{V(H$\alpha$)\tablenotemark{a}}& \colhead{V(Na\,\sc{i})\tablenotemark{b}}&\colhead{$\sigma$(H$\alpha$)\tablenotemark{c}}\\
\colhead{} & \colhead{(km s$^{-1}$)} & \colhead{(km s$^{-1}$)}&\colhead{(km s$^{-1}$)}}
\startdata
K1  & -56 & -29: & 61  \\
K2  & -32 & $-$ & 58  \\
K3  & -65 & 30  & 80  \\
K4  & 13  & 31  & 68  \\
K5  & 54  & 101 & 35  \\
K6  & 54  & 115 & 20  \\
K7  & 36  & 78  & 30  \\
K8  & 84  & 42  & 70  \\
K9  & 20  & -6:  & 65  \\
K10 & 54  & 86:  & 100 \\
K11 & -2  & 26  & 19  \\
K12 & -78 & $-$ & 36  
\enddata
\tablecomments{\footnotesize The mean errors for the velocity measurements in H$\alpha$ and Na\,{\sc{i}} for this bundle are 27 km s$^{-1}$ and 60 km s$^{-1}$ respectively.'':'' means large uncertainties. Velocities referred to nucleus (region A) of galaxy IC~694. (a) V$_{A}$(H$\alpha$)=3121$\pm$27\,km\,s$^{-1}$.(b) V$_{A}$(Na\,{\sc{i}})=3057$\pm$34\,km\,s$^{-1}$. (c) Relative error in $\sigma$ is $\simeq$13\%}\label{kinearp299}
\end{deluxetable}                                                         

\clearpage

\begin{deluxetable}{lccc}
\tabletypesize{\footnotesize}
\tablewidth{0pc}
\tablecolumns{4}
\tablecaption{Kinematic of IC~694 regions}
\tablehead{\colhead{Region} & \colhead{V(H$\alpha$)\tablenotemark{a}}& \colhead{V(Na\,\sc{i})\tablenotemark{b}}&\colhead{$\sigma$(H$\alpha$)\tablenotemark{c}}\\
\colhead{} & \colhead{(km s$^{-1}$)} & \colhead{(km s$^{-1}$)}&\colhead{(km s$^{-1}$)}}
\startdata
A  &0    &0    &108\\
A1 &-86  &-47  &73 \\
A2 &-7   &22   &64 \\
A3 &-134 &-86  &36 \\
A4 &-151 &-96  &50 \\
A5 &-67  &-25  &45 \\
A6 &-44  &-37  &40 \\
A7 &-57  &-26  &41 \\
F  &-163 &$-$  &72 \\
G  &-121 &-77  &43 \\
H1 &-130 &-89  &61 \\
H2 &-117 &-93  &109\\
I  &60   &98   &61 \\
I1 &37   &$-$  &79 
\enddata
\vspace{-0.4cm}
\tablecomments{\footnotesize Velocities referred to nucleus (region A). (a) V$_{A}$(H$\alpha$)=3121$\pm$27\,km\,s$^{-1}$.(b) V$_{A}$(Na\,{\sc{i}})=3057$\pm$34\,km\,s$^{-1}$. (c) Relative error in $\sigma$ is $\simeq$10\%}\label{kineic694}
                               
\end{deluxetable}                                                           

\clearpage

\begin{deluxetable}{lccc}
\tabletypesize{\footnotesize}
\tablewidth{0pc}
\tablecolumns{4}
\tablecaption{Kinematic of NGC~3690 regions}
\tablehead{\colhead{Region} & \colhead{V(H$\alpha$)\tablenotemark{a}}& \colhead{V(Na\,\sc{i})\tablenotemark{b}}&\colhead{$\sigma$(H$\alpha$)\tablenotemark{c}}\\
\colhead{} & \colhead{(km s$^{-1}$)} & \colhead{(km s$^{-1}$)}&\colhead{(km s$^{-1}$)}}
\startdata
B1      & 0    &0    &172 \\
B2      & -116 &-55  &110 \\
B11     & 38   &50   &143 \\
B16     & -46  &-15  &117 \\
B21     & -18  &9    &128 \\
B22     & -121 &-48  &78  \\
B25     & -112 &-64  &54  \\
C1      & 32   &65   &93  \\
C2      & 16   &51   &99  \\
C3      & 13   &36   &90  \\
C4      & -24  &8    &94  \\
C6      & 115  &$-$  &82  \\
C$^{'}$ & 101  &$-$  &69  \\
D3      & -101 &-40  &67  \\
D4      & -112 &-52  &54  
\enddata
\tablecomments{\footnotesize Velocities referred to nucleus (regions B1). (a) V$_{B1}$(H$\alpha$)=3040$\pm$27\,km\,s$^{-1}$.(b) V$_{B1}$(Na\,{\sc{i}})=2976$\pm$32\,km\,s$^{-1}$. (c) Relative error in $\sigma$ is $\simeq$10\%}\label{kinengc3690} 
                               
\end{deluxetable}                                                           

\clearpage 

\begin{deluxetable}{lcc}
\tabletypesize{\footnotesize}
\tablewidth{0pc}
\tablecolumns{5}
\tablecaption{Dynamical mass estimates}
\tablehead{\colhead{Region} & \colhead{R$_{eff}$\tablenotemark{a}} &
  \colhead{M$_{dyn}$\tablenotemark{b}}\\
\colhead{} & \colhead{(kpc)} & \colhead{($\times$10$^9$M$_{\odot}$)}}
\startdata
A       & 0.054$-$0.062 & 1.10$-$1.3 \\
B1      & 0.035$-$0.022 & 1.81$-$1.1\\
B2      & 0.030$-$0.025 & 0.63$-$0.5\\
C1      & 0.022$-$0.042 & 0.33$-$0.6\\
C$^{'}$ & 0.023$-$0.043 & 0.19$-$0.3\\
\enddata
\tablecomments{Dynamical mass estimates for the nuclei of the system (A and B1) and main H$\alpha$-selected emitting regions (B2, C1 and C$'$). The stellar and gas effective radius have been derived for the HST F110M (first column) and for F190N (Pa$\alpha$) (second column). In both cases the dynamical mass has been derived using the H$\alpha$ velocity dispersion values. See \S \ref{dynmass} for details.}\label{tablamasic}
\end{deluxetable}                             

\clearpage

\clearpage

\end{document}